\documentstyle[preprint,aps]{revtex}

\begin{document}
\draft

\title{Anomalous temperature dependence of resistivity in
  quasi-one-dimensional conductors in a strong magnetic field}

\author{Anatoley T. Zheleznyak\cite{Zheleznyak} and
Victor M. Yakovenko\cite{Yakovenko}}
\address{Department of Physics and Center for Superconductivity
Research, University of Maryland, College Park,
MD 20742-4111}
\date{v.3: 26 Mar 1999 (v.1: 16 Feb 1998), cond-mat/9802172, 
Eur. Phys. J. B 11, 385--399 (1999)}
\maketitle
\begin{abstract}
  We present a heuristic, semiphenomenological model of the anomalous
  temperature (T) dependence of resistivity $\rho_{xx}$ recently
  observed experimentally in the quasi-one-dimensional (Q1D) organic
  conductors of the $\rm(TMTSF)_2X$ family in moderately strong
  magnetic fields. We suggest that a Q1D conductor behaves like an
  insulator ($d\rho_{xx}/dT<0$), when its effective dimensionality is
  one, and like a metal ($d\rho_{xx}/dT>0$), when its effective
  dimensionality is greater than one. Applying a magnetic field
  reduces the effective dimensionality of the system and switches the
  temperature dependence of resistivity between the insulating and
  metallic laws depending on the magnitude and orientation of the
  magnetic field. We critically analyze whether various microscopic
  models suggested in literature can produce such a behavior and find
  that none of the models is fully satisfactory. In particular, we
  perform detailed analytical and numerical calculations within the
  scenario of magnetic-field-induced spin-density-wave precursor
  effect suggested by Gor'kov and find that the theoretical results do
  not agree with the experimental observations.
\end{abstract}


\section{Experimental Introduction}
\label{sec:intro}

In recent experiments \cite{Cooper94,Jerome95,Jerome96b,Chaikin98}, a
very unusual temperature ($T$) dependence of resistivity was observed
in quasi-one-dimensional (Q1D) organic conductors $\rm(TMTSF)_2ClO_4$
(at the ambient pressure) and $\rm(TMTSF)_2PF_6$ (at a pressure about
9 kbar) \cite{TMTSF} in moderately strong magnetic fields $H$ of the
order of 10 T at $T\lesssim50$ K.  Unexpectedly large
magnetoresistance in these materials has already attracted attention
in earlier measurements \cite{Jacobsen81,Cooper84,Cooper86}.

The $\rm(TMTSF)_2X$ materials consist of one-dimensional (1D)
conducting chains parallel to the crystal axis {\bf a} (for general
reviews of the $\rm(TMTSF)_2X$ materials, see Refs.\
\cite{Cooper94,Jerome94,Yamaji90,Gorkov84,Jerome82}). The chains are
weakly coupled in the two others directions {\bf b} and {\bf c}, the
coupling in the {\bf c} direction being much weaker than in the {\bf
b} direction. In zero magnetic field or in the field parallel to the
{\bf b} axis, the resistivity of $\rm(TMTSF)_2PF_6$ along the chains,
$\rho_{aa}$, depends on temperature approximately quadratically:
$\rho_{aa} \sim T^2$ \cite{Cooper94,Chaikin98,Jerome98}, which is
consistent with the standard Fermi-liquid theory \cite{AGD}, provided
resistivity is dominated by electron-electron scattering. When a
magnetic field is applied along the {\bf c} axis, $\rho_{aa}$ does not
change appreciably at high temperatures; however, below a certain
magnetic-field-dependent temperature $T_{\rm min}\sim 20$ K,
resistivity starts to grow with decreasing temperature:
$d\rho_{aa}/dT<0$ at $T<T_{\rm min}$
\cite{Cooper94,Jerome96b,Chaikin98}. In other words, the behavior of
the system changes from metallic, $d\rho_{aa}/dT>0$, to insulating,
$d\rho_{aa}/dT<0$, when the temperature is lowered below $T_{\rm
min}$.  The temperature $T_{\rm min}$ increases with the increase of
magnetic field. Such a behavior is very surprising in view of the fact
that no thermodynamic phase transition is observed in the system at
this relatively high temperature $T_{\rm min}\sim 20$ K. (A phase
transition into the magnetic-field-induced spin-density-wave (FISDW)
state takes place at the transition temperature $T_c\sim 2$ K, which
is an order of magnitude lower than $T_{\rm min}$.) As the temperature
is lowered further below $T_{\rm min}$, $\rho_{aa}(T)$ continues to
grow until another temperature scale $T_{\rm max}\sim 8\:{\rm
K}<T_{\rm min}$ is reached. Magnetoresistance is huge at $T\sim T_{\rm
max}$: $\rho_{aa}$ at $H=7.8$ T is about 10 times greater than
$\rho_{aa}$ at $H=0$ \cite{Chaikin98}. At the lower temperatures
$T<T_{\rm max}$, behavior of the system starts to depend crucially on
the exact orientation of the magnetic field
\cite{Chaikin98,Jerome96-note}. If the magnetic field lies in a plane
formed by the direction {\bf a} of the chains and another integer
crystallographic direction, such as {\bf c} or ${\bf c}+{\bf b}$,
$\rho_{aa}$ recovers the metallic behavior $d\rho_{aa}/dT>0$ at
$T<T_{\rm max}$. For other, generic orientations of the magnetic
field, $\rho_{aa}$ retains the nonmetallic behavior, either continuing
to grow with decreasing temperature: $d\rho_{aa}/dT<0$, or saturating
at a high constant value. The temperature $T_{\rm max}$ does not
depend appreciably on the magnetic field. If at $T<T_{\rm max}$ the
magnetic field is rotated in the plane perpendicular to the direction
{\bf a} of the chains, $\rho_{aa}$ exhibits sharp minima when the
field is aligned with the integer planes described above (the
so-called ``magic angles'' effect). This effect was discovered
experimentally earlier \cite{Osada91,Naughton91,Chaikin92d} following
the theoretical suggestion by Lebed' \cite{Lebed86,Lebed89a}. Finally,
at low temperatures, the system may enter the magnetic-field-induced
spin-density-wave (FISDW) phase at the transition temperature
$T_c\sim\mbox{1--2}$ K, where $\rho_{aa}$ increases sharply. Other
diagonal components of the resistivity tensor of $\rm(TMTSF)_2PF_6$
seem to behave similarly to $\rho_{aa}$: The angular dependence of
$\rho_{cc}$ \cite{Chaikin98,Chaikin94b} at $T<T_{\rm max}$ and the
temperature dependences of $\rho_{cc}$ \cite{Balicas97,Balicas} and
$\rho_{bb}$ \cite{Balicas,Chashechkina,Moser} appear to be
qualitatively similar to those of $\rho_{aa}$.  However, in the very
recent measurements \cite{Chashechkina} it was found that, when the
magnetic field is oriented along the magic direction {\bf c},
$\rho_{cc}$ monotoniuosly decreases with temperature, whereas
$\rho_{aa}$ and $\rho_{bb}$ exhibit a minimum at $T=T_{\rm min}$.
When the magnetic field is tilted away from {\bf c}, $\rho_{cc}$
develops a minimum at a temperature much lower than $T_{\rm min}$ for
$\rho_{aa}$ and $\rho_{bb}$.

In $\rm(TMTSF)_2ClO_4$, a magnetic field also causes $\rho_{aa}(T)$,
as well as the NMR relaxation rate $1/T_1$, to grow with decreasing
temperature at $T<T_{\rm min}$ \cite{Jerome95,Jerome95-note}, which is
an indication of a nonmetallic behavior induced by the magnetic field.
On the other hand, it was shown in Refs.\ 
\cite{Chaikin94b,Chaikin97a,Chaikin97b} that the angular and temperature
dependences of $\rho_{cc}$ in $\rm(TMTSF)_2ClO_4$ are quite different
from those in $\rm(TMTSF)_2PF_6$ and can be interpreted within the
standard Fermi-liquid picture. The behavior of $\rm(TMTSF)_2ClO_4$ may
or may not \cite{Chaikin97b} be complicated by doubling of the crystal
period in the {\bf b} direction occurring in this material at $T=24$
K.  To avoid complications, we will focus our theoretical study on
$\rm(TMTSF)_2PF_6$, which has a simple crystal structure.

The behavior of resistivity in $\rm(TMTSF)_2PF_6$ completely
contradicts the conventional Fermi-liquid picture of a metal with an
open Fermi surface. In this picture, applying a magnetic field
perpendicular to the direction of the chains should produce no or very
little magnetoresistance, should not alter the metallic temperature
dependence of resistance, and should exhibit no magic angles effect
\cite{Zheleznyak95a}. Thus, unconventional theoretical approaches are
required.

\section{Heuristic Theoretical Picture}
\label{sec:qualitative}

We suggest that the following theoretical picture may qualitatively
account for the unusual behavior of $\rm(TMTSF)_2PF_6$
\cite{conferences}.

It is well known theoretically (see, for example, Ref.\ 
\cite{Yakovenko87b}) that the orbital effect of a {\bf c}-axis
magnetic field $H$ on a Q1D conductor is characterized by the
cyclotron energy $E_H=ebHv_F/c$, where $e$ is the electron charge, $b$
is the distance between the chains in the {\bf b} direction, $v_F$ is
the Fermi velocity, and $c$ is the speed of light.  For the realistic
values of the model parameters (see Sec.\ \ref{sec:H}), we estimate
that $E_H/H\approx1.8$ K/T. The magnitude of the cyclotron energy,
$E_H\approx14$ K at $H=7.8$ T, is close to the temperature of the
resistivity minimum at that magnetic field, $T_{\rm min}\sim20$ K.
Taking into account that the minimum of resistivity clearly has a
magnetic origin (it does not exist without magnetic field), and
$T_{\rm min}$ grows with the increase of magnetic field, we suggest
that the minimum of resistivity occurs when the temperature reaches
the energy scale of the magnetic field; that is, $T_{\rm min}\approx
E_H$.

Now we need to identify the nature of the second energy scale in the
problem, the temperature of the resistivity maximum, $T_{\rm max}$. At
the temperatures $T>T_{\rm max}$, it appears that $\rho_{aa}$ depends
only on the magnetic field projection on the ${\bf c}^*$ axis
perpendicular to the {\bf a} and {\bf b} directions \cite{Hcos}. From
this observation, we may conclude that at $T>T_{\rm max}$ the system
behaves effectively as a two-dimensional (2D) system; that is, the
coupling between the chains in the {\bf c} direction is not relevant. On
the other hand, at $T<T_{\rm max}$ the coupling along the {\bf c} axis
becomes important. This is manifested by the magic angles effect, which
is an essentially three-dimensional (3D) phenomenon involving both the
{\bf b} and {\bf c} axes. The coupling between the chains along the {\bf
c} axis is characterized by the electron tunneling amplitude $t_c$,
whose magnitude is believed to be of the order of 10 K \cite{Grant83},
which is close to $T_{\rm max}\sim8$ K. Thus, we suggest that the
electron tunneling amplitude $t_c$ sets the temperature scale $T_{\rm
max}$ of the resistivity maximum: $T_{\rm max}\approx t_c$. This
conjecture is supported by the experimental fact that $T_{\rm max}$
(unlike $T_{\rm min}$) does not depend appreciably on the magnetic field
\cite{Jerome96b,Chaikin98}. We also need to mention that, according to
Refs.\ \cite{Chaikin95,Jacobsen83}, the coupling between the chains in
the {\bf b} direction, $t_b$, is much greater than the temperatures
discussed in our paper: $t_b\sim300$ K.

Taking into account these energy scales, we identify three qualitatively
different regimes in the behavior of a Q1D system in a magnetic field:

1) High temperatures: $E_H\approx T_{\rm min}<T<t_b$. In this region,
the temperature is greater than both the magnetic energy $E_H$ and the
electron tunneling amplitude $t_c$ along the {\bf c} axis, but lower
than the tunneling amplitude $t_b$ along the {\bf b} axis. Thus, we
may neglect both the magnetic field and the coupling between the
chains along the {\bf c} axis and treat the system as a normal 2D
Fermi liquid without magnetic field. This results in the quadratic law
$\rho_{aa} \sim T^2$ and the metallic behavior of the resistivity
$d\rho_{aa}/dT>0$.

2) Intermediate temperatures: $t_c\approx T_{\rm max}<T<T_{\rm
min}\approx E_H$. In this region, the temperature is still greater
than the coupling between the chains along the {\bf c} axis, so the
system remains 2D; however, the effect of the magnetic field becomes
important. It is known that, in the presence of a magnetic field along
the {\bf c} axis, the motion of electrons along the {\bf b} axis
becomes quantized, and the dispersion law of electrons becomes
one-dimensional (1D) \cite{Lebed84}. The degeneracy of the electron
spectrum in the {\bf b} direction is a specific manifestation of the
Landau degeneracy in a magnetic field in the case of a 2D system with
a strong Q1D anisotropy. This phenomenon is called
``one-dimensionalization'' of a Q1D system by a magnetic field
\cite{Gorkov84}. Even though the spectrum of electrons becomes 1D and
their wave functions become localized in the {\bf b} direction, the
wave functions still spread over many chains (if $E_H\ll t_b$), which
results in a considerable interaction between different chains
\cite{Yakovenko87b}. So the system is not truly 1D, because it does
not consist of completely decoupled 1D chains. Nevertheless, we may
expect that, at least, some 1D features would be present in this
regime and, via a mechanism that need to be specified, would lead to
an insulating transport behavior $d\rho_{aa}/dT<0$. In general, 1D
systems have stronger tendency toward insulating behavior than
higher-dimensional systems, because various insulating mechanisms,
such as renormalization of umklapp scattering, density-wave
instabilities, and Anderson localization, are more effective in one
dimension than in higher dimensions. So the conjecture that the
insulating behavior is caused by the magnetic-field-enforced
``one-dimensionalization'' is plausible, but requires detailed
studying of a specific mechanism \cite{Cooper94}. We review possible
candidates for the mechanism in the next section and quantitatively
analyze one of the mechanisms in rest of the paper.

3) Low temperatures: $T<T_{\rm max}\approx t_c$. In this region, the
coupling between the $({\bf a},{\bf b})$ planes becomes important. The
magnetic field pointing exactly along the {\bf c} axis does not affect
the electron motion along that axis. Thus, in addition to the
magnetic-field-enforced 1D dispersion law discussed in part 2), the
system acquired an extra dispersion in the {\bf c} direction and
becomes effectively 2D, which results in a metallic, Fermi-liquid
behavior $d\rho_{aa}/dT>0$. If the magnetic field does not point along
the {\bf c} axis, the component of the field perpendicular to the {\bf
  c} axis suppresses the energy dispersion along that axis, so the
system remains effectively 1D and insulating: $d\rho_{aa}/dT<0$. If
the direction of the field is close to the {\bf c} axis, we expect
resistivity to decrease with decreasing temperature in the range
$E^{(c)}_H<T<T_{\rm max}\approx t_c$ and to start increasing again at
$T<E^{(c)}_H$, where $E^{(c)}_H$ is the cyclotron energy of the
electron motion along the {\bf c} axis, which is proportional to the
projection of the magnetic field perpendicular to the {\bf c} axis.
The same arguments apply not only to the {\bf c} axis, but also to the
${\bf c}+{\bf b}$ axis and other integer crystallographic directions
$m{\bf c}+n{\bf b}$. However, because the electron tunneling
amplitudes in these directions decrease rapidly with the increase of
the integers $m$ and $n$, the effect is clearly visible experimentally
only for the ${\bf c}+{\bf b}$ axis.

In summary, we suggest that the unusual transport behavior of
$\rm(TMTSF)_2PF_6$ results from the changes in the effective
dimensionality of the system caused by the applied magnetic field. The
system is 2D at $E_H<T<t_b$ and effectively 1D at $t_c<T<E_H$. At
$T<t_c$ the system is effectively 2D for the magic orientations of the
magnetic field and effectively 1D for generic orientations. Whenever the
system is 2D (or 3D), it is a normal Fermi liquid, and the temperature
dependence of resistivity is metallic. Whenever the system is
effectively 1D, the temperature dependence of resistivity is insulating.
The latter state of the system might be called the
magnetic-field-induced Luttinger insulator (MFILI), by analogy with the
term ``Luttinger liquid'', which refers to the metallic state of a 1D
system \cite{Haldane81}.

We do not have detailed mathematical calculations that can prove the
heuristic picture outlined in this section. Nevertheless, we can
predict some experimental effects based on this picture. In Refs.\ 
\cite{Chaikin94a,Chaikin95}, oscillations of $\rho_{cc}$ upon rotation
of a magnetic field in the $({\bf a},{\bf c})$ plane were discovered
by Danner {\it et al.} Following the theoretical suggestion of Ref.\ 
\cite{Anderson94b}, it was found that a small magnetic field along the
{\bf b} axis destroys the oscillations \cite{Chaikin95}. We predict
that if a magnetic field is rotated in the magic plane from the ${\bf
  b}+{\bf c}$ direction toward the {\bf a} direction, the Danner
oscillations should exist, even though the magnetic field has a finite
{\bf b}-component. The suggested geometry has an advantage over the
geometry of experiment \cite{Chaikin95}, where the magnetic field had
a fixed {\bf b}-component, that the Danner oscillations would not be
mixed up with the Lebed' oscillations occurring when the magnetic
field is rotated in the $({\bf b},{\bf c})$ plane. This prediction is
based on the idea that the Danner oscillations require that the
electron motion in the third direction is not suppressed by the
magnetic field, which happens only when the magnetic field belongs to
a magic plane.  We also predict that the Danner oscillations should
disappear at $T>T_{\rm max}$, where the electron dispersion in the
third direction is smeared out by temperature.  A detailed study of
magnetic oscillations upon rotation of the magnetic field around the
$\bf c^*$ axis with different tilts relative to the $\bf c^*$ axis was
recently performed by Lee and Naughton \cite{Naughton98}.  They found
possible to interprete most, but not all, of the results within a
conventional semiclassical theory of metals.

\section{Review of Theoretical Models}
\label{sec:review}

Within the heuristic framework presented in the previous Section, a
theoretical study of the problem reduces to the following two parts:

(a) How a magnetic field induces the negative temperature dependence
of resistivity $d\rho_{aa}/dT<0$ in a 2D metal with a strong Q1D
anisotropy.  (2D problem)

(b) How the electron tunneling in the third direction does or does not
suppress the effect found in part (a) depending on the orientation of
the magnetic field in the $({\bf b},{\bf c})$ plane. (3D problem)

From the above formulation, it is clear that the 3D problem (b) can be
addressed only after the 2D problem (a) has been solved. In other words,
we believe that the insulating temperature dependence of resistivity
(problem (a)) and the drop of resistivity at the magic angles (problem
(b)) have a common origin.

However, until recently, theoretical and experimental efforts were
focused on solving problem (b) without recognizing and addressing
problem (a). Soon after the experimental discovery of the magic angles
\cite{Osada91,Naughton91,Chaikin92d}, a number of theories tried to
explain this effect semiclassically
\cite{Maki92b,Osada92a,Chaikin92c}.  The theories
\cite{Maki92b,Osada92a} found the magic angles effect in $\rho_{bb}$
and $\rho_{cc}$, but not in $\rho_{aa}$. Microscopic analysis
\cite{Zheleznyak95a} of the ``hot spots'' model \cite{Chaikin92c}
demonstrated that it cannot explain the huge magnetoresistance and the
magic angles effect. In all of these models, resistivity was
calculated by introducing a relaxation time $\tau$ phenomenologically
and studying semiclassical electron trajectories in the momentum
space. These theories assumed that relaxation mechanism does not
change dramatically as the magnetic field is rotated. In view of
experiment \cite{Chaikin98}, this assumption is completely wrong,
because it is the switching between metallic and insulating states and
the corresponding change in the relaxation mechanism that causes the
magic angles effect.

Another explanation of the magic angles was proposed on the basis of
the Luttinger liquid concept in Ref.\ \cite{Anderson94b} (see also
Ref.\ \cite{Chaikin}). This theory suggested that the magic angles
effect reflects the change of the effective dimensionality of the
system: The system is a 2D Luttinger liquid at generic angles, and a
3D normal Fermi liquid at the magic angles. The prediction of the
theory that even a small {\bf b}-component of the magnetic field would
destroy coherence of the interchain hopping in the {\bf c} direction
was confirmed experimentally in Ref.\ \cite{Chaikin95,Chaikin}.
However, this theory focuses only on problem (b), but does not address
the issue of the temperature dependence of resistivity and does not
explain how problem (a) may be solved. Moreover, in the actual
calculations \cite{Anderson94b}, $t_c$ is treated as a perturbation to
a 1D system, and the much greater tunneling amplitude $t_b$ is
effectively neglected.  For this reason, the theory \cite{Anderson94b}
actually studies the dimensionality crossover between 2D and 1D, not
2D and 3D, and the very important {\bf c}-component of the magnetic
field does not appear in these calculations.

The very first calculation of the angular dependence of $\rho_{aa}$
was done by Lebed' and Bak \cite{Lebed89a} before the experimental
discovery of the magic angles. However, it predicted maxima, not
minima, of resistance at the magic angles. This discrepancy was
corrected in the subsequent work \cite{Lebed94}. In this theory,
resistivity $\rho_{aa}\propto1/\tau$ is studied by calculating the
rate of umklapp scattering $1/\tau$ in the lowest order of
perturbation theory. This approach allows to study how resistivity
depends on temperature and on the magnitude and orientation of the
magnetic field. According to Ref.\ \cite{Lebed94}, the temperature
dependence of the scattering rate changes with the magnetic field
orientation: $1/\tau \propto T^2$ at the magic angles and $1/\tau
\propto T$ at generic angles, because the electron dispersion law is
2D and 1D in these cases, respectively. At low enough temperatures,
the difference between the $T^2$ and $T$ laws should result in sharp
dips of resistance at the magic angles. However, the theory predicts
that both temperature dependences are metallic ($d\rho_{aa}/dT>0$),
whereas experimentally the temperature dependence is insulating
($d\rho_{aa}/dT<0$) for nonmagic angles at $T<T_{\rm min}$ and for all
angles at $T_{\rm max}<T<T_{\rm min}$
\cite{Jerome95,Jerome96b,Chaikin98}. Thus, the theory \cite{Lebed94} is
not adequate either.

In Ref.\ \cite{Lebed89a}, Lebed' and Bak considered also the
scattering rate of electrons on impurities renormalized by the
electron-electron interaction in the lowest order and noted that it
grows with decreasing temperature, when a magnetic field is applied.
It is easy to check analytically that this scattering rate is
approximately constant without magnetic field and, if $t_c$ is
neglected, grows as $\ln(E_H/T)$ in a magnetic field at $T<E_H$. We
calculated this diagram numerically in the 2D case and found that it
increases by only about 20\% in the relevant range of fields and
temperatures, which is insufficient to explain the experiment. (Our
numerical calculation of the umklapp electron-electron scattering
diagram in the 2D case also confirmed that application of a magnetic
field changes the temperature dependence of $1/\tau$ from $T^2$ to $T$
at $T<E_H$, as discussed in the preceding paragraph.)

After the magnetic-field-induced insulating temperature dependence of
$\rho_{aa}$ was discovered experimentally, theory started to address
this problem specifically. Refs.\ \cite{Jerome95,Jerome96b} suggested
that resistance increases at $T<T_{\rm min}$, because the
``one-dimensionalization'' of $\rm(TMTSF)_2X$ by a magnetic field induces
formation of a pseudogap in the charge channel. It is well known that
charge and spin excitations are independent in a 1D system
\cite{Jerome82}. In the presence of umklapp scattering and repulsive
interactions, charge excitations may develop a pseudogap, whereas spin
excitations may remain gapless. This results in insulating temperature
dependence of resistivity coexisting with metallic behavior of spin
susceptibility. This effect is observed experimentally in the
sulfur-based compounds $\rm(TMTTF)_2X$ \cite{Coulon82a}, which are
more 1D than the selenium-based compounds $\rm(TMTSF)_2X$ (see Refs.\ 
\cite{Jerome96b,Jerome94,Jerome98}).  Induction of a charge pseudogap
by a magnetic field would explain insulating behavior of all
components of the resistivity tensor in $\rm(TMTSF)_2PF_6$
\cite{Balicas97,Balicas,Chashechkina,Moser} and the NMR data in
$\rm(TMTSF)_2ClO_4$ \cite{Jerome95} (although the transverse
resistivity in $\rm(TMTSF)_2ClO_4$ behaves differently
\cite{Chaikin97a,Chaikin97b}).

However, in order to achieve quantitative agreement with the
experiment, this theory assumes that $t_b\sim50$ K \cite{Jerome95} or
15--30 K \cite{Jerome96b}, which is almost an order of magnitude
smaller that the commonly accepted value $t_b\approx250$ K deduced
from the {\bf b}-axis plasma edge \cite{Jacobsen83} and the ({\bf
  a},{\bf b})-plane angular magnetic oscillation \cite{Chaikin95}. In
this theory, the magnitude of the pseudogap is determined by the
energy scale where the renormalization-group (RG) equations (also
called the parquet equations \cite{Dzyaloshinskii72a}) for the
forward, backward, and umklapp scattering amplitudes of
electron-electron interaction diverge. However, because the chains are
strongly coupled, it is not correct to limit the RG equations to only
those three amplitudes of interaction. It is necessary to include an
infinite number of the amplitudes of interaction between different
chains, which makes the RG equations integro-differential
\cite{Gorkov74b}. For a Q1D conductor in a magnetic field, the
integro-differential RG equations were derived and solved numerically
without umklapp in Refs.\ 
\cite{Yakovenko87b,Yakovenko87a,Yakovenko91a}. The temperature where a
solution of the RG equations diverges was interpreted in the latter
papers as the FISDW transition temperature $T_c$. On the other hand,
in a purely 1D case, the energy where a solution of the RG equations
diverges is conventionally interpreted as a pseudogap energy, not as a
transition temperature, because thermodynamic phase transitions are
not possible in 1D systems \cite{Landau-V}. Thus, the problem is
whether the RG equations for a Q1D conductor in a magnetic field can
simultaneously describe formation of a pseudogap at a high temperature
and the FISDW transition at a much lower temperature. In a purely 1D
case this is possible, because the RG equations separate exactly into
two independent sets of equations for the spin and charge channels
\cite{Dzyaloshinskii72a}. However, in a higher-dimensional case, all
of the interaction amplitudes are coupled, and separation of the RG
equations into independent channels does not seem feasible. It is not
clear why a pseudogap energy and a FISDW transition temperature would
differ by an order of magnitude in the RG approach, if both are
produced by the same mechanism of ``one-dimensionalization'' and enhancement
of the Peierls susceptibility by a magnetic field.

Even after the RG equations are solved, the temperature dependence of
resistivity still needs to be calculated. This issue was addressed in
the theory by Gor'kov \cite{Gorkov95,Gorkov96}, which suggested that
resistivity is given by the same diagram that was studied in Refs.\ 
\cite{Lebed89a,Lebed94}, but with a renormalized,
temperature-dependent umklapp scattering amplitude. In the standard
model of the FISDW transition \cite{Montambaux85}, where $t_b\gg E_H$,
and electron-electron interaction is repulsive, it is safe to neglect
the superconducting channel contribution to the RG equations
\cite{Yakovenko87b}. In this case, the RG equations reduce to the
conventional ladder-RPA equations. The umklapp scattering amplitude,
renormalized according to these equations, grows when $T\to T_c$.
Because $\rho_{aa}$ is proportional to the square of the umklapp
amplitude, the insulating regime $d\rho_{aa}/dT<0$ might be achieved
if the umklapp amplitude grows faster than the phase space factor $T$
or $T^2$ decreases. One might expect, though, that resistivity would
start to grow only in a narrow vicinity of the FISDW transition, not
at the temperatures an order of magnitude higher than $T_c$. Whether
this approach can quantitatively explain the experiment, particularly
the role of a magnetic field, can be verified by numerical
calculations. In the rest of the paper, we study this issue in detail.
The approach of Refs.\ \cite{Gorkov95,Gorkov96} is attractive, because
it permits straightforward calculation of transport coefficients and
connects naturally with the standard, successful model of FISDW on one
hand and with the simpler, better-understood transport model of Refs.\ 
\cite{Lebed89a,Lebed94} on the other hand.

\section{Q1D conductor in a magnetic field}
\label{sec:Q1D}

For the following theoretical description, we select the $x$, $y$, and
$z$ axes along the crystal directions {\bf a}, {\bf b}, and {\bf c},
which are not orthogonal in the triclinic (TMTSF)$_2$X crystals.
Electrons can tunnel between different chains with the amplitudes
$t_{\bf l}$, where ${\bf l}=(l_1,l_2)$ is a 2D integer vector that
determines the transverse displacement of the electron, ${\bf d_l} =
l_1{\bf b} + l_2{\bf c}$, in the process of tunneling. The Fermi
surface of 1D electron motion along the chains consists of two Fermi
points characterized by the Fermi wave vectors $\pm k_F$. We label the
electrons with the wave vectors close to $\pm k_F$ by the index
$\alpha=\pm$. In the vicinity of the Fermi energy, the energy
dispersion law of the longitudinal electron motion can be linearized
as $\varepsilon=\pm\hbar v_F k_x$, where $\hbar$ is the Planck
constant, $v_F$ is the Fermi velocity, the energy $\varepsilon$ is
counted from the Fermi energy, and the longitudinal wave vector $k_x$
is counted from $\pm k_F$ for the two Fermi points. In the absence of
magnetic field, the total, longitudinal and transverse, electron
dispersion law is
\begin{equation}
\varepsilon_\alpha({\bf k})=\alpha\hbar v_F k_x
+\sum_{\bf l}2t_{\bf l}\cos({\bf k}_\perp {\bf d_l}
+\alpha\varphi_{\bf l}),
\label{dispersion} 
\end{equation}
where ${\bf k}=(k_x,k_y,k_z)=(k_x,{\bf k}_\perp)$ are the electron wave
vectors along the $x$, $y$, and $z$ axes. The phases $\varphi_{\bf l}$
in the transverse dispersion law (\ref{dispersion}) of a triclinic
crystal are determined by the amplitudes of tunneling to different
molecules belonging to the same chain \cite{Yamaji86}. In this paper, we
present analytical formulas for a general dispersion law
(\ref{dispersion}) with any number of the transverse tunneling
amplitudes $t_{\bf l}$, but we perform numerical calculations only for
the 2D case with the two tunneling amplitudes: $t_b=t_{1,0}$ between the
nearest and $t_b'=t_{2,0}$ between the next-nearest chains in the {\bf
b} direction. The values of the corresponding phases, $\varphi_b$ and
$\varphi_b'$, are not known reliably. According to Yamaji
\cite{Yamaji86}, $\varphi_b'=-\pi/2$, and $\varphi_b$ varies from
$7^\circ$ to $40^\circ$ when temperature varies from 300 K to 1.7 K in
$\rm(TMTSF)_2PF_6$. In our numerical calculations, we assume
$\varphi_b'=2\varphi_b$ for simplicity and consider several values of
$\varphi_b$ between the two extremal values $\varphi_b=0$ and
$\varphi_b=\pi/4$.

Suppose that a magnetic field {\bf H} is applied perpendicular to the
chains. It can be introduced into the Hamiltonian of the system via
the Peierls-Onsager substitution ${\bf k}\rightarrow{\bf k}-x[{\bf
  H}\times{\bf e}_x]e/\hbar C$, where ${\bf e}_x$ is the unit vector
along the $x$ axis, $e$ is the electron charge, and $C$ is the speed
of light.  The eigenfunctions $\psi_{\alpha, {\bf k}}$ of
noninteracting electrons in the magnetic field $H$ are
\cite{Yakovenko87b}
\begin{equation}
   \psi_{\alpha, {\bf k}} (x, n_y, n_z)
   =\frac{1}{\sqrt{L\cal N}}\exp 
   \left[ i \left(k_xx + k_y n_y b + k_z n_z c
   +\alpha\sum_{\bf l}\frac{2t_{\bf l}}{\hbar v_F G_{\bf l}} 
   \sin({\bf k}_\perp{\bf d_l}-G_{\bf l}x+\alpha\varphi_{\bf l})
   \right) \right],
\label{wave_funct}
\end{equation}
where $G_{\bf l}={\bf e}_x\cdot[{\bf d_l}\times{\bf H}]e/\hbar C$ are
the wave vectors proportional to the magnetic field, $n_y$ and $n_z$ are
the integer coordinates of the chains in the $y$ and $z$ directions, $L$
is the length of a chain, and $\cal N$ is the total number of the
chains. The eigenenergies of eigenfunctions (\ref{wave_funct}) are
\begin{equation}
\varepsilon = \alpha \hbar v_F k_x,
\label{dispersion1D}
\end{equation}
thus the electron dispersion law is effectively 1D in the magnetic
field. 

The single-particle Green function of noninteracting electrons in
the magnetic field $H$ was found in Ref.\ \cite{Lebed84}:
\begin{eqnarray}
&&{\cal G}_{\alpha}(x,x',{\bf k}_{\perp},i\omega_m)=
  \int_{-\infty}^{\infty} \frac{dk_x}{2\pi} 
  \frac{\exp[ik_x(x-x')]}
  {i\omega_m-\alpha\hbar v_F k_x}  \nonumber \\ 
&&\times \exp\left(
  i\alpha\sum_{\bf l}\frac{2t_{\bf l}}{\hbar v_FG_{\bf l}} 
  [\sin({\bf k}_{\perp}{\bf d_l}-G_{\bf l}x+\alpha\varphi_{\bf l})
  -\sin({\bf k}_{\perp}{\bf d_l}-G_{\bf l}x'+\alpha\varphi_{\bf l})]
  \right),
\label{GrFunct}
\end{eqnarray}
where $\omega_m=2\pi(m+1/2)T$ is the Matsubara frequency. The Green
function (\ref{GrFunct}) is a product of two terms: The first term is
the Green function of 1D electrons, whereas the second, exponential
term represents the transverse motion of the electrons. Only the
second term contains the magnetic field via the wave vectors $G_{\bf
  l}$, which introduce periodic dependences on $x$ and make the Green
function (\ref{GrFunct}) not translationally invariant along the
chains. In 2D case, we denote the magnetic wave vector $G_{1,0}$ as
simply $G$. The integral over $k_x$ in Eq.\ (\ref{GrFunct}) can be
easily taken, but the Lehmann representation (\ref{GrFunct}) is more
convenient for analytic continuation from the Matsubara frequencies to
the real frequencies.

\section{Renormalization of the umklapp scattering amplitude due to
SDW instability}
\label{sec:g3}

The tendency of a Q1D system toward the SDW instability manifests
itself in divergence of the density-wave susceptibility, which is
shown diagrammatically in the lowest order in Fig.\ \ref{fig:DWloop}.
In this figure, the solid and dashed lines represent the Green
functions of the $+$ and $-$ electrons, and $\Omega_m$ and ${\bf
  q_{\perp}}$ are the incoming energy and the transverse wave vector.
In the Matsubara technique, the analytic expression for the bare
susceptibility per one chain is \cite{Lebed84}
 \begin{eqnarray}
     \chi_0(x'-x,{\bf q}_{\perp},i\Omega_m) & = &
      -T \sum_j \int \frac{bc\,d^2{\bf k}_\perp}{(2\pi)^2}
      G_-(x,x',{\bf k}_{\perp}+{\bf q}_{\perp},i\omega_j+i\Omega_m)
      G_+(x',x,{\bf k}_{\perp},i\omega_j) \nonumber \\
       & = &\frac{T}{2 \pi (\hbar v_F)^2} 
      \frac{\lambda(x'-x,{\bf q}_{\perp})}{\sinh[2\pi T(x'-x)/\hbar v_F]}
      \int_{-\infty}^{\infty}
      \frac{d\varpi \sin[(x'-x)\varpi/ \hbar v_F]}{\varpi-i\Omega_m},
\label{chi}
\end{eqnarray}
where
\begin{equation}
    \lambda(x,{\bf q}_\perp)=\int\frac{bc\,d{\bf k}_\perp}{(2\pi)^2} 
    \exp\left(-i\sum_{\bf l}\frac{8t_{\bf l}}{\hbar v_F G_{\bf l}}
    \sin(G_{\bf l}x/2)\cos({\bf k}_\perp{\bf d_l})
    \cos({\bf q}_\perp{\bf d_l}/2-\varphi_{\bf l}) \right).
\label{lambda} 
\end{equation}
In Eqs.\ (\ref{chi}) and (\ref{lambda}) and elsewhere, the integration
over the transverse wave vectors ${\bf k}_\perp$ goes over the Brillouin
zone. Because of the averaging over ${\bf k}_{\perp}$, susceptibility
(\ref{chi}) is a translationally invariant function of $x'-x$, unlike
the Green function (\ref{GrFunct}). This allows us to Fourier-transform
Eq.\ (\ref{chi}) over $x'-x$ and to obtain $\chi_0({\bf q},i\Omega_m)$ as a
function of the 3D wave vector ${\bf q}$. As follows from Eq.\
(\ref{chi}), $\chi_0(x,{\bf q}_{\perp},i\Omega_m)$ behaves as $1/x$ when
$x\leq v_F/2\pi T$. This results in logarithmical divergence of
$\chi_0({\bf q},i\Omega_m)$, which we cut off at a small distance
$x_0=1/2\gamma k_F$ \cite{Montambaux85}, where $\gamma$ is the Euler
constant \cite{cutoff}. 

Because the electron conduction band in $\rm(TMTSF)_2X$ is
half-filled, the Fermi wave vector is commensurate with the crystal
lattice wave vector along the chains: $4k_F=2\pi/a$. This relation
permits the umklapp scattering process, where two ``+'' electrons are
transformed into two ``--'' electrons, and the change of the total
electron wave vector, $4k_F$, is absorbed into the lattice wave vector
$2\pi/a$. The amplitude of this process is conventionally denoted by
$\gamma_3$ \cite{Dzyaloshinskii72a}. The one-loop diagram of Fig.\ 
\ref{fig:DWloop} generates a ladder renormalization of the vertices of
interaction between electrons as shown diagrammatically in Fig.\ 
\ref{fig:vertices}.  In this figure, the wavy lines represent the
bare, unrenormalized vertices of forward ($g_2$) and umklapp ($g_3$)
scattering, the circles represent the corresponding renormalized
vertices $\gamma_2$ and $\gamma_3$, and the thin lines inside the
circles indicate spin conservations along the electron lines.
Appropriate for the SDW channel, the + and $-$ electron lines in Fig.\ 
\ref{fig:vertices} have opposite spins. The equations of Fig.\ 
\ref{fig:vertices} are the same as in Ref.\ {\protect
  \cite{Gorkov96}}. The analytic expression of these equations is
\begin{eqnarray}
   \gamma_2({\bf q}, i\Omega_m) = g_2 + 
    g_2 \chi_0({\bf q},i\Omega_m) \gamma_2({\bf q},i\Omega_m) +
    g_3 \chi_0(-{\bf q},-i\Omega_m) \gamma_3({\bf q},i\Omega_m), 
\label{gamma2eq} \\
   \gamma_3({\bf q}, i\Omega_m) = g_3 + 
    g_2 \chi_0(-{\bf q},-i\Omega_m) \gamma_3({\bf q},i\Omega_m) +
    g_3 \chi_0({\bf q},i\Omega_m) \gamma_2({\bf q},i\Omega_m),
\label{gamma3eq}
\end{eqnarray}
and the solution is
\begin{eqnarray}
   \gamma_2({\bf q}, i\Omega_m) =
   \frac{g_2-(g_2^2-g_3^2)\chi_0(-{\bf q}, -i\Omega_m)}
   {[1-g_2\chi_0({\bf q},i\Omega_m)] [1-g_2\chi_0(-{\bf q},-i\Omega_m)]
   -g_3^2\chi_0({\bf q}, i\Omega_m) \chi_0(-{\bf q},-i\Omega_m)}, 
\label{gamma2} \\
   \gamma_3({\bf q}, i\Omega_m) = \frac{g_3} 
   {[1-g_2\chi_0({\bf q},i\Omega_m)] [1-g_2\chi_0(-{\bf q},-i\Omega_m)]
   -g_3^2\chi_0({\bf q}, i\Omega_m) \chi_0(-{\bf q},-i\Omega_m)}.
\label{gamma3} 
\end{eqnarray}
Notice that whenever the umklapp vertex appears in the r.h.s.\
of Eqs.\ (\ref{gamma2eq}) and (\ref{gamma3eq}), the signs of ${\bf q}$
and $i\Omega_m$ in the argument of $\chi_0$ are reversed.

As temperature decreases, the bare susceptibility $\chi_0({\bf q},0)$
grows until the denominator in Eqs.\ (\ref{gamma2}) and (\ref{gamma3})
vanishes at a certain temperature $T_c$, which is the FISDW transition
temperature:
\begin{equation}
   [1-g_2\chi_0({\bf q},0)] [1-g_2\chi_0(-{\bf q},0)]
   -g_3^2\chi_0({\bf q},0) \chi_0(-{\bf q},0) = 0.
\label{condFISDW}  
\end{equation}
The interaction amplitudes $\gamma_{2,3}({\bf q},0)$ diverge at the
transition temperature. Since Eq.\ (\ref{condFISDW}) is a quadratic form
of $\chi_0({\bf q},0)$ and $\chi_0(-{\bf q},0)$, it vanishes at two
different temperatures for a given value of {\bf q}. Usually, it is
assumed that only the higher temperature is physically significant, and
the transition temperature is determined from Eq.\ (\ref{condFISDW}) by
selecting the wave vector {\bf q} that provides the maximal value for
$T_c$. 
 
By considering equations similar to Eq.\ (\ref{condFISDW}), Lebed'
\cite{LebedG3} predicted that the umklapp splitting of the FISDW
instability would result in oscillations of $T_c$ vs $H$, but the
effect was not observed experimentally. On the other hand, the
experimental $T$-$H$ phase diagram of $\rm(TMTSF)_2PF_6$ can be well
reproduced while neglecting umklapp \cite{Montambaux85}. In order not
to spoil the phase diagram by the umklapp splitting and to avoid
unnecessary complications, we assume that $g_3$ is sufficiently small
and neglect it in Eq.\ (\ref{condFISDW}) and in the denominator of
Eq.\ (\ref{gamma3}) \cite{Dupuis}.  Using the conventional
band-structure parameters of $\rm(TMTSF)_2PF_6$: $t_a=2900$ K
($E_F=\sqrt{2}t_a$), $t_b=t_a/10$, and $v_F=2\times10^5$ m/s, we find
that the choice $t_b'=20$ K and $\tilde{g}_2=g_2/2\pi\hbar v_F=0.2288$
produces a $T$-$H$ phase diagram close to the one observed
experimentally in Ref.\ \cite{Chaikin92b} at 12 kbar. The phase
diagram does not depend on $\varphi_b$ and is shown in Fig.\ 
\ref{fig:PhaseDiagram}. In this figure, various symbols indicate the
integer number $N$ of the quantized longitudinal wave vector of FISDW:
$q_x=NG$ \cite{Montambaux85}. Note that wide spacing in $H$ between
the points of our calculations would not allow us to observe the
oscillations of $T_c$ vs $H$ \cite{LebedG3}, even if we took into
account the umklapp splitting of the FISDW instability. If we set
$t_b'=0$, the electron spectrum (\ref{dispersion}) acquires perfect
nesting:
$\varepsilon_+(k_x,k_y)=-\varepsilon_-(k_x,k_y-(\pi-2\varphi_b)/b)$,
and the SDW transition temperature $T_0=(\hbar v_F /\pi x_0)
\exp(-1/\tilde{g}_2)=14.7$ K becomes independent of the magnetic
field.

While the Matsubara representation of $\gamma_{2,3}$ is useful for
determining the $T$-$H$ phase diagram, we need the scattering vertices
at the real frequencies to calculate resistivity. Analytically
continuing Eq.\ (\ref{chi}) from the Matsubara frequencies $i\Omega_m$
to the real energies $\varepsilon$, we find the following expression
for the bare susceptibility:
\begin{equation}
    \chi_0({\bf q},\varepsilon) =  
    \frac{T}{(\hbar v_F)^2} \int_{x_0}^{\infty} dx \,
    \frac{{\rm Re}\left\{\exp(-iq_xx) \lambda(x,{\bf q}_{\perp})\right\} 
    \exp(ix\varepsilon/\hbar v_F)}
    {\sinh(2\pi T x /\hbar v_F)}.
\label{chi:Re.freq}
\end{equation}
Substituting Eq.\ (\ref{chi:Re.freq}) into Eqs.\ (\ref{gamma2}) and
(\ref{gamma3}), we find the scattering vertices $\gamma_{2,3}({\bf
q},\varepsilon)$ at the real energies $\varepsilon$.

\section{Umklapp resistivity of a Q1D metal}
\label{sec:resistivity}

Using the variational principle for the Boltzmann equation \cite{Ziman},
we find the following expression for the resistivity along the chains
due to electron-electron umklapp scattering:
\begin{equation}
   \rho_{xx}=\frac{2(\pi\hbar)^2 bc}{e^2 T L \cal N}
   \sum_{{\bf k}_1,{\bf k}_2,{\bf k}_3,{\bf k}_4}
   f_0(\varepsilon_1) f_0(\varepsilon_2)
   [1-f_0(\varepsilon_3)][1-f_0(\varepsilon_4)] \,
   W_{{\bf k}_1,{\bf k}_2}^{{\bf k}_3,{\bf k}_4},
\label{rho:var.pr}
\end{equation}
where $\varepsilon_i = \varepsilon({\bf k}_i)$ are the energies of
electrons in the eigenstates $|{\bf k}_i\rangle$ (\ref{wave_funct}),
$f_0(\varepsilon)$ is the Fermi distribution function, and $W_{{\bf
k}_1,{\bf k}_2}^{{\bf k}_3,{\bf k}_4}$ is the scattering rate of two
electrons from the states $|{\bf k}_1\rangle$ and $|{\bf k}_2\rangle$
into the states $|{\bf k}_3\rangle$ and $|{\bf k}_4\rangle$:
\begin{equation}
   W_{{\bf k}_1,{\bf k}_2}^{{\bf k}_3,{\bf k}_4}=\frac{2\pi}{\hbar}\,
   |\langle{\bf k}_1,{\bf k}_2|\gamma_3|{\bf k}_3,{\bf k}_4\rangle|^2\,
   \delta(\varepsilon_1+\varepsilon_2-\varepsilon_3-\varepsilon_4).
\label{Tr.prob}
\end{equation}
Here $\delta(\varepsilon)$ is the Dirac delta
function, and the matrix element is
\begin{eqnarray}
&&  \langle{\bf k}_1,{\bf k}_2|\gamma_3|{\bf k}_3,{\bf k}_4\rangle =
    \sum_{n_y^{(1)}, n_y^{(2)}, n_z^{(1)}, n_z^{(2)}}
    \int_{-\infty}^{\infty} dx_1 \, dx_2 \,
    \psi_{+, {\bf k}_1} (x_1, n_y^{(1)}, n_z^{(1)})\,
    \psi_{+, {\bf k}_2} (x_2, n_y^{(2)}, n_z^{(2)}) \nonumber \\
&&  \times\psi_{-, {\bf k}_3}^* (x_2, n_y^{(2)}, n_z^{(2)})\,
    \psi_{-, {\bf k}_4}^* (x_1, n_y^{(1)}, n_z^{(1)})\,
    \gamma_3 (x_1-x_2, n_y^{(1)}-n_y^{(2)}, n_z^{(1)}-n_z^{(2)},      
    \varepsilon_2-\varepsilon_3).
\label{matrix.el}
\end{eqnarray}
The vertex of interaction $\gamma_3$ is written in the mixed, coordinate
and energy, representation, describing the umklapp scattering of the two
``+'' electrons with the coordinates $x_1$ and $x_2$ located on the
chains $(n_1^{(y)},n_1^{(z)})$ and $(n_2^{(y)},n_2^{(z)})$ from the
states $|{\bf k}_1\rangle$ and $|{\bf k}_2\rangle$ into the two ``$-$''
states $|{\bf k}_3\rangle$ and $|{\bf k}_4\rangle$ with the same
coordinates.

Substituting Eq.\ (\ref{wave_funct}) into Eq.\ (\ref{matrix.el}) and
changing the variables of integration $k_x^{(i)} \to \varepsilon_i$ via
Eq.\ (\ref{dispersion1D}), we find the following expression for the
resistivity
\begin{equation}
   \rho_{xx}=\frac{2(\pi\hbar)^2 L^3 bc}{e^2 T \cal N} 
   \int_{-\infty}^\infty 
   \frac{d\varepsilon_1\, d\varepsilon_2\, d\varepsilon_3\, d\varepsilon_4}
   {(2\pi \hbar v_F)^4}  f_0(\varepsilon_1) f_0(\varepsilon_2)
   [1-f_0(\varepsilon_3)] [1-f_0(\varepsilon_4)] \,
   W_{\varepsilon_1, \varepsilon_2}^{\varepsilon_3, \varepsilon_4}, 
\label{rho:var.pr.Q1D}
\end{equation}
where
\begin{eqnarray}
&& W_{\varepsilon_1, \varepsilon_2}^{\varepsilon_3, \varepsilon_4} =
   \sum_{{\bf k}_{\perp}^{(1)},{\bf k}_{\perp}^{(2)},{\bf k}_{\perp}^{(3)},
   {\bf k}_{\perp}^{(4)}} W_{{\bf k}_1, {\bf k}_2}^{{\bf k}_3, {\bf k}_4}
\nonumber \\
&& =\frac{bc \cal N}{\hbar L^3} 
   \int_{-\infty}^{\infty} dx'\,dx''\int\frac{d^3{\bf q}}{(2\pi)^2}
   \,|\gamma_3({\bf q}, \varepsilon_2-\varepsilon_3)|^2\,
   \delta(\varepsilon_1+\varepsilon_2-\varepsilon_3-\varepsilon_4)
\nonumber \\   
&& \times\exp\left[ i\left(q_x
   +\frac{\varepsilon_1+\varepsilon_4}{\hbar v_F}\right)x'\right]
   \lambda(x', -{\bf q}_{\perp})
   \exp\left[ i\left(-q_x
   +\frac{\varepsilon_2+\varepsilon_3}{\hbar v_F}\right)x''\right]
   \lambda(x'', {\bf q}_{\perp}).
\label{Tr.prob.Q1D}
\end{eqnarray}
Using the $\delta$-function from Eq.\ (\ref{Tr.prob.Q1D}), we take the
integral over $\varepsilon_4$ in Eq.\ (\ref{rho:var.pr.Q1D}). Then,
changing the integration variables $\varepsilon_1$, $\varepsilon_2$, and
$\varepsilon_3$ to $\varepsilon = \varepsilon_2 - \varepsilon_3$,
$\varepsilon_1'= \varepsilon_1 + (\varepsilon_2-\varepsilon_3)/2$, and
$\varepsilon_2'= \varepsilon_2 + \varepsilon_3$ and taking the integrals
over $\varepsilon_1'$ and $\varepsilon_2'$, we obtain the final
expression for resistivity:
\begin{equation}
   \rho_{xx} = \frac{\hbar (bc)^2}{32 e^2 T} 
   \int_{-\infty}^{\infty}d\varepsilon\,
   \int d^3 {\bf q}\:
   F({\bf q},\varepsilon)\, F(-{\bf q},\varepsilon)\,
   |{\tilde \gamma}_3({\bf q},\varepsilon)|^2,
\label{rho_xx.Q1D}        
\end{equation}
where
\begin{eqnarray}
&& F({\bf q}, \varepsilon)=\frac{2T}{\hbar v_F \sinh(\varepsilon /2T)}
   \int_{-\infty}^\infty dx \,
   \frac{\exp(-iq_x x) \lambda(x,{\bf q}_{\perp}) 
   \sin(x\varepsilon/\hbar v_F)}{\sinh(2\pi T x /\hbar v_F)}, 
\label{Def:F} \\
&& {\tilde \gamma}_3({\bf q},\varepsilon)=
   \frac{\gamma_3({\bf q},\varepsilon)}{2\pi \hbar v_F},
\label{gamma:tilde}
\end{eqnarray}
are dimensionless functions.  The renormalized umklapp vertex
$\gamma_3({\bf q},\varepsilon)$, given by Eqs.\ (\ref{gamma3}) and
(\ref{chi:Re.freq}), should be substituted into Eqs.\ 
(\ref{rho_xx.Q1D}) and (\ref{gamma:tilde}). If the renormalization of
$\gamma_3$ is neglected ($\gamma_3=g_3$), the integral over
$\varepsilon$ in Eq.\ (\ref{rho_xx.Q1D}) can be taken analytically,
and the result agrees with the expressions obtained in Refs.\ 
\cite{Lebed89a,Lebed94}.

\section{Temperature dependence of resistivity in zero magnetic field}
\label{sec:H=0}

Having obtained the general expression (\ref{rho_xx.Q1D}) for the
longitudinal resistivity, let us examine the limit of zero magnetic
field first. In this case, $F({\bf q},\varepsilon)$ (\ref{Def:F})
becomes
\begin{equation}
   F({\bf q},\varepsilon)=\frac{bc}{(2\pi)^2}\int
   \frac{d^2{\bf k}_{\perp}}
   { \cosh \left(\frac{\textstyle \varepsilon}{\textstyle 2T}\right)+
   \cosh\left(\frac{\textstyle  \hbar v_F q_x+
   \sum_{\bf l}4t_{\bf l}\cos({\bf q}_\perp {\bf d_l}/2-\varphi_{\bf l})
   \cos({\bf k}_{\perp}{\bf d_l})} {\textstyle 2T} \right) }.
\label{F(H=0)}
\end{equation}
The function $F({\bf q},\varepsilon)$ (\ref{F(H=0)}), confines
integration in Eq.\ (\ref{rho_xx.Q1D}) to the energy interval
\begin{equation}
|\varepsilon|\alt T
\label{E<T}
\end{equation}
and to the wave-vectors region defined by the two inequalities:
\begin{eqnarray}
&&  |\hbar v_F q_x + \sum_{\bf l} 4t_{\bf l}
    \cos ({\bf q}_\perp{\bf d_l}/2 - \varphi_{\bf l})
    \cos({\bf k}_\perp{\bf d_l})| \alt T,
\label{constraint+} \\
&&  |\hbar v_F q_x - \sum_{\bf l} 4t_{\bf l}
    \cos ({\bf q}_\perp{\bf d_l}/2 + \varphi_{\bf l})
    \cos({\bf k}_\perp'{\bf d_l})| \alt T.
\label{constraint-} 
\end{eqnarray}
It follows from Eqs.\ (\ref{constraint+}) and (\ref{constraint-}) that
the integration over the transverse wave vectors is restricted by the
inequality
\begin{equation}
   |\Xi({\bf q_\perp,k_\perp,k_\perp'})| \alt T,
\label{Sigma<T} 
\end{equation}
where
\begin{equation}
    \Xi({\bf q_\perp,k_\perp,k_\perp'})
    =\sum_{\bf l}4t_{\bf l} [\cos({\bf q}_\perp{\bf d_l}/2
    - \varphi_{\bf l}) \cos({\bf k}_\perp{\bf d_l})
    +\cos ({\bf q}_\perp{\bf d_l}/2 + \varphi_{\bf l})
    \cos({\bf k}_\perp'{\bf d_l})].
\label{Sigma} 
\end{equation}

The integrals over $\varepsilon$ and $q_x$ in Eq.\ (\ref{rho_xx.Q1D})
with the function $F({\bf q},\varepsilon)$ (\ref{F(H=0)}) can be taken
analytically, provided we neglect dependence of $\gamma_3(q_x,{\bf
q}_\perp,\varepsilon)$ on $q_x$ and $\varepsilon$, that is, replace
$\gamma_3(q_x,{\bf q}_\perp,\varepsilon)$ by $\gamma_3(\bar{q}_x,{\bf
q}_\perp,\bar{\varepsilon})$, where $\bar{q}_x$ and $\bar{\varepsilon}$
are some characteristic values of $q_x$ and $\varepsilon$ from the
intervals of integration (\ref{E<T}), (\ref{constraint+}), and
(\ref{constraint-}):
\begin{equation}
  \rho_{xx} \approx \frac{\pi^2 (bc)^4 T}{16 e^2 v_F}
  \int\frac{d^2{\bf q}_\perp d^2{\bf k}_\perp d^2{\bf q}_\perp'}
  {(2\pi)^6}
  \left(\frac{\Xi({\bf q_\perp,k_\perp,k_\perp'})/T}
  {\sinh[\Xi({\bf q_\perp,k_\perp,k_\perp'})/4T]}\right)^2
  |{\tilde \gamma}_3(\bar{q}_x,{\bf q}_\perp,\bar{\varepsilon})|^2.
\label{rho_xx.unid} 
\end{equation}
This approximation should be valid provided the peak in
$\gamma_3(q_x,{\bf q}_\perp,\varepsilon)$ at $\varepsilon=0$ and {\bf q}
equal to the nesting vector is wider than temperature, which is the case
for temperatures not very close to the transition temperature $T_c$.
Eq.\ (\ref{rho_xx.unid}) with $\Xi$ (\ref{Sigma}) coincides with the
expression obtained by Gor'kov \cite{Gorkov95,Gorkov96} for
$\varphi_{\bf l}=0$, except that we find an additional factor $1/2$ in
the argument of $\sinh$, as in Ref.\ \cite{Gorkov97}.

If renormalization of $\gamma_3$ is neglected, and $\gamma_3$ is
replaced by a constant $g_3$, then integral (\ref{rho_xx.unid}) gives
the volume of the wave-vectors space restricted by inequality
(\ref{Sigma<T}). In a general 2D or 3D case, this volume is proportional
to $T$, thus resistance (\ref{rho_xx.unid}) is proportional to $T^2$:
\begin{equation}
   \rho_{xx}^{(\rm 2D,3D)} \propto {\tilde g}_3^2\,T^2,
\label{rho_xx.23D} 
\end{equation}
which is the standard result of the Fermi-liquid theory \cite{AGD}. In
1D case \cite{resistivity}, where $\Xi=0$, Eq.\ (\ref{rho_xx.unid})
reproduces the result of Ref.\ \cite{Gorkov73}:
\begin{equation}
   \rho_{xx}^{(\rm 1D)} \approx \frac{\pi^2 bc T}{e^2 v_F}\,
   |{\tilde \gamma}_3^{(\rm 1D)}(\bar{q}_x,\bar{\varepsilon})|^2.
\label{rho_xx.1D} 
\end{equation}
The 1D resistance (\ref{rho_xx.1D}) is proportional to temperature
$T$ multiplied by the square of the renormalized umklapp amplitude
$\gamma_3$, which may also depend on temperature.

Now let us consider the simplest 2D case where only one tunneling
amplitude $t_b$ is kept in Eq.\ (\ref{Sigma}). If $\varphi_b=0$ (similar
equations hold also for $\varphi_b =\pi/2$), then Eq.\ (\ref{Sigma}) can
be factorized \cite{Gorkov95,Gorkov96}:
\begin{equation}
    \Xi(q_y,k_y,k_y')\Big|_{\varphi_b=0}
    =8t_b \cos(q_yb/2) \cos[(k_y+k_y')b/2] \cos[(k_y-k_y')b/2].
\label{Sigma0} 
\end{equation}
If renormalization of $\gamma_3$ is neglected ($\gamma_3=g_3$), then
Eq.\ (\ref{rho_xx.unid}) with $\Xi$ from Eq.\ (\ref{Sigma0}) gives:
\begin{equation}
   \rho_{xx}^{(\rm 2D)}\Big|_{\varphi_b=0}
   \propto {\tilde g}_3^2\,T^2\ln^2(t_b/T),
\label{rho_xx.2D} 
\end{equation}
which has an extra logarithmic factor compared to Eq.\ 
(\ref{rho_xx.23D}) \cite{Gorkov97}. Eqs.\ (\ref{rho_xx.23D}) and
(\ref{rho_xx.2D}) are in agreement with the results of Ref.\ 
\cite{Zheleznyak95a}, where the so-called ``hot spots'' in the
distribution of the umklapp scattering time over the Fermi surface of
a Q1D metal were studied. The ``hot spots'' are the points where the
scattering rate is strongly enhanced compared to the rest of the Fermi
surface, typically by the factor $\ln(t_b/T)$ and occasionally by the
factor $\sqrt[3]{t_b/T}$.  Positions of the hot spots are determined
by the saddle points of the function $\Xi(q_y,k_y,k_y')$
\cite{Zheleznyak95a}. When $\varphi_b\neq0,\pi/2$, only isolated hot
spots exist on the Fermi surface. Because they occupy a small phase
space, the hot spots do not contribute significantly to resistivity in
this general case, and Eq.\ (\ref{rho_xx.23D}) holds. However, in the
special case $\varphi_b=0,\pi/2$, the entire Fermi surface becomes
``hot'' \cite{Zheleznyak95a}, and resistivity acquires the logarithmic
factor of Eq.\ (\ref{rho_xx.2D}).

In the special 2D case with only one tunneling amplitude $t_b$ and
$\varphi_b=0$, the electron dispersion (\ref{dispersion}) has a perfect
nesting at the wave vector $q_x=0$ and $q_y=\pi/b$. The zero-field
susceptibility $\chi_0({\bf q},\varepsilon)$, given by Eqs.\
(\ref{chi:Re.freq}) and (\ref{lambda}), diverges logarithmically at
$\varepsilon=q_x=0$ and $q_y=\pi/b$. Since we neglect $g_3$ in the
denominator of Eq.\ (\ref{gamma3}), the renormalized vertex
$\gamma_3(0,\pi/b,0)$ becomes
\begin{equation}
   \gamma_3(0,\pi/b,0)\propto\frac{g_3}{g_2^2\,\ln^2(T/T_0)},
\label{gamma3T0}  
\end{equation}
where $T_0=\hbar v_F /(\pi x_0) \exp(-1/\tilde{g}_2)$ is the SDW
transition temperature. At the same time, condition (\ref{Sigma<T}) with
$\Xi$ given by Eq.\ (\ref{Sigma0}) restricts integration in Eq.\
(\ref{rho_xx.unid}) to the vicinity of either $q_y=\pi/b$ or
$k_y+k_y'=\pi/b$ or $k_y-k_y'=\pi/b$. The first of these conditions is
satisfied at the same wave vector $q_y=\pi/b$ where $\gamma_3$
(\ref{gamma3T0}) diverges. Assuming that the integral in Eq.\
(\ref{rho_xx.unid}) is dominated by the vicinity of $q_y=\pi/b$, we find:
\begin{equation}
   \rho_{xx}^{(\rm 2D)}\Big|_{\varphi_b=0}
   \propto {\tilde \gamma}_3^2(0,\pi/b,0)\,T^2\ln^2(t_b/T)
   \propto\frac{{\tilde g}_3^2\,T^2\ln^2(t_b/T)}
   {{\tilde g}_2^4\,\ln^4(T/T_0)}.
\label{rho_xx.T0} 
\end{equation}
Eq.\ (\ref{rho_xx.T0}) is analogous to the expression obtained by
Gor'kov \cite{Gorkov95,Gorkov96}, but differs in the powers of the
logarithms.  The factor $T^2$ in Eq.\ (\ref{rho_xx.T0}) tends to
diminish resistivity with decreasing temperature, which is
characteristic for a metal. On the other hand, the logarithmic factors
in Eq.\ (\ref{rho_xx.T0}), both in the numerator and denominator, tend
to increase resistance, which is characteristic for an insulator.
Which of these two competing tendencies wins can be found numerically.

In Fig.\ \ref{fig:rxx}, we show temperature dependences of resistivity
$\rho_{xx}(T)$ calculated via Eqs.\ (\ref{rho_xx.Q1D}) and
(\ref{F(H=0)}) in 2D case at zero magnetic field. Because we neglected
$g_3$ in the denominator of Eq.\ (\ref{gamma3}), $\rho_{xx}\propto
g_3^2$ exactly. In Figs.\ \ref{fig:rxx}, \ref{fig:rxxH}, and
\ref{fig:rxxH30/40}, we plot the ratio $\rho_{xx}/\tilde{g}_3^2$ where
the dimensionless umklapp scattering amplitude
$\tilde{g}_3=g_3/2\pi\hbar v_F$ cancels out. As discussed in Appendix,
the value of $\tilde{g}_3$ can be recovered by comparing these figures
with the experimental data.  The top solid curve in Fig.\ 
\ref{fig:rxx} shows $\rho_{xx}(T)$ calculated with only one tunneling
amplitude $t_b$ and $\varphi_b=0$. We observe that $\rho_{xx}(T)$ has
a positive, metallic slope $d\rho_{xx}(T)/dT>0$ at high temperatures,
whereas at lower temperatures the slope is negative. It is surprising
that the negative slope starts at the temperature about 60 K, which is
several times higher than the SDW transition temperature $T_0=14.7$ K.
This indicates that the logarithmic factors in Eq.\ (\ref{rho_xx.T0})
overcome the $T^2$ factor at relatively high temperatures.

Now let us discuss a more general 2D model with only one tunneling
amplitude $t_b$, but with $\varphi_b\neq0$. In this case, $\gamma_3$
diverges at the nesting vector $q_x=0$ and $q_y=(\pi-2\varphi_b)/b$. The
umklapp vertex $\gamma_3(0,(\pi-2\varphi_b)/b,0)\propto g_3/\ln(T/T_0)$
diverges at exactly the same transition temperature $T_0$ as in the case
$\varphi_b=0$, but less strongly than in Eq.\ (\ref{gamma3T0}), because
$\chi_0(0,q_y,0)$ and $\chi_0(0,-q_y,0)$ in Eq.\ (\ref{gamma3}) do not
diverge at the same wave vector $q_y$. At $\varphi_b\neq0$, condition
(\ref{Sigma<T}) with $\Xi$ given by Eq.\ (\ref{Sigma}) is not
satisfied at the nesting vector $q_y=(\pi-2\varphi_b)/b$ independently
of $k_y$ and $k_y'$, which further reduces $\rho_{xx}(T)$ compared to
Eq.\ (\ref{rho_xx.T0}) for $\varphi_b=0$. Temperature dependences
$\rho_{xx}(T)$ are shown in Fig.\ \ref{fig:rxx} by solid curves for
different values of $\varphi_b$. All curves have the same values of
$t_b=290$ K, $\tilde{g}_2 = 0.2288$, and $T_0=14.7$ K as the top solid
curve with $\varphi_b=0$. While all solid curves diverge at the same
transition temperature $T_0$, the region of the negative slope in
$\rho_{xx}(T)$ shrinks rapidly with increasing $\varphi_b$ and becomes
much smaller than $T_0$ at $\varphi_b\agt\pi/27$. Thus, a non-zero phase
$\varphi_b$ strongly suppresses the precursor effect in resistivity. For
the curves with $\varphi_b\agt\pi/27$, the behavior of $\rho_{xx}(T)$
qualitatively follows the 2D quadratic law (\ref{rho_xx.23D}) at the low
temperatures $T \alt 2t_b/\pi=185$ K and the 1D law (\ref{rho_xx.1D}) at
the higher temperatures $T \agt 2t_b/\pi$. The slope of $\rho_{xx}(T)$
at $T \agt 2t_b/\pi$ is rather small, presumably because the 1D
logarithmic renormalization of $\gamma_3$ partially compensates the
linear temperature factor in Eq.\ (\ref{rho_xx.1D}).

The dotted curves in Fig.\ \ref{fig:rxx} represent $\rho_{xx}(T)$ for
the 2D model where the tunneling amplitude to the
next-nearest-neighboring chains, $t_b'=20$ K, is introduced in addition
to $t_b$. Since $t_b'$ eliminates nesting in the dispersion law
(\ref{dispersion}), the system does not have SDW instability provided
$t_b'>T_0$, so $\gamma_3$ does not diverge. The slope of the curve with
$\varphi_b=0$ changes from negative to positive at $T \alt 2t_b'/\pi$,
which creates a maximum in the $\rho_{xx}(T)$ curve. Thus, $t_b'$ cuts
off the precursor effect in $\rho_{xx}(T)$ at $T\sim t_b'/\pi$, as
discussed in Refs.\ \cite{Gorkov95,Gorkov96}.

In conclusion, we have confirmed the suggestion by Gor'kov
\cite{Gorkov95,Gorkov96} that, in zero magnetic field, the
renormalization of umklapp amplitude due to proximity to a SDW
transition can produce a negative slope in $\rho_{xx}(T)$ at
temperatures much higher than the SDW transition temperature $T_0$.
This precursor effect is suppressed by a non-zero phase $\varphi_b$ in
the electron dispersion law, which does not influence the SDW
transition temperature, but shrinks the temperature region of the
negative slope. The second tunneling amplitude $t_b'$ suppresses the
SDW transition temperature and cuts off the negative slope at $T\alt
t_b'/\pi$. In $\rm(TMTSF)_2X$, negative slope of $\rho_{xx}(T)$ is not
observed in zero magnetic field and appears only when a magnetic field
is applied. According to the scenario suggested by Gor'kov
\cite{Gorkov95,Gorkov96}, the negative slope is eliminated by a
non-zero $t_b'$ (and, possibly, by a non-zero $\varphi_b$, as shown
above), but it is restored by a magnetic field. We examine feasibility
of this scenario in the next section.

\section{Temperature dependence of resistivity in a magnetic field}
\label{sec:H}

In 2D case, the effect of a magnetic field on a Q1D electron system is
characterized by the wave vector $G=ebH/\hbar c$ and the cyclotron
energy $E_H=\hbar v_F G=ebHv_F/c$ \cite{b}. Using the parameters
$v_F=2\times10^5$ m/s and $b=7.7$ \AA, we find that $E_H/H\approx1.8$
K/T. Magnetic field enters Eqs.\ (\ref{rho_xx.Q1D}) and (\ref{Def:F})
for resistivity only through the function $\lambda(x,q_y)$
(\ref{lambda}). In 2D case, magnetic field makes $\lambda$ a periodic
function of $x$ with the period $2\pi/G$:
$\lambda(x,q_y)=\lambda(x+2\pi/G,q_y)$, so it can be expanded into a
Fourier series:
\begin{equation}
  \lambda(x,q_y) = \sum_{n=-\infty}^{\infty} A_n(q_y)\,e^{inGx},
\label{lambda:FT}
\end{equation}
with some coefficients $A_n$. Substituting Eq.\ (\ref{lambda:FT}) into
Eq.\ (\ref{Def:F}), we find:
\begin{equation}
  F({\bf q},\varepsilon) = \sum_{n=-\infty}^{\infty} \frac{A_n(q_y)}
  {\cosh \left(\frac{\textstyle \varepsilon}{\textstyle 2T}\right)+
  \cosh\left(\frac{\textstyle \hbar v_F (q_x-nG)}
  {\textstyle 2T} \right) }.
\label{F:H}
\end{equation}
As follows from Eq.\ (\ref{F:H}), the integration in Eq.\
(\ref{rho_xx.Q1D}) is concentrated in the energy interval
$|\varepsilon|\alt T$ and, at $T\ll E_H$, in the vicinity of the integer
wave vectors $q_x\approx nG$. Assuming that the characteristic width in
$q_x$ and $\varepsilon$ of the function $\gamma_3({\bf q},\varepsilon)$
is greater than temperature, one could replace $\gamma_3({\bf q},
\varepsilon)$ by $\gamma_3(nG, q_y,0)$ in the integral
(\ref{rho_xx.Q1D}). In this case, the integrals over $\varepsilon$ and
$q_x$ can be taken analytically:
\begin{eqnarray}
\rho_{xx}^{(2D)}&\approx&\frac{\pi^2 b^2c T}{e^2 v_F} 
   \int \, dq_y \sum_{n=-\infty}^{\infty} 
   |{\tilde \gamma}_3(nG,q_y,0)|^2 \nonumber \\
&& \times\sum_{n_1, n_2=-\infty}^{\infty} A_{n_1}(q_y)A_{n_2}(-q_y)
   \left(\frac{(n_1+n_2-n)E_H/4T}
   {\sinh[(n_1+n_2-n)E_H/4T]} \right)^2.
\label{rho_xx.Q1D.H}
\end{eqnarray} 
At temperatures much lower than the cyclotron energy, $T\ll E_H$, only
the term with $n_1+n_2=n$ contributes significantly to Eq.\
(\ref{rho_xx.Q1D.H}):
\begin{equation}
   \rho_{xx}^{(2D)} \approx \frac{\pi^2 b^2c T}{e^2 v_F}
   \int \, dq_y \sum_{n=-\infty}^{\infty} 
   |{\tilde \gamma}_3(nG,q_y,0)|^2
   \sum_{n_1=-\infty}^{\infty} A_{n_1}(q_y)A_{n-n_1}(-q_y).
\label{rho_xx.Q1D.H>T}
\end{equation}
Eq.\ (\ref{rho_xx.Q1D.H>T}) is similar to the 1D formula
(\ref{rho_xx.1D}), except for the additional integration over $q_y$ and
summation over $n$. This is a consequence of one-dimensionalization of the
electron spectrum (\ref{dispersion1D}) by the magnetic field.

If renormalization of $\gamma_3$ is neglected ($\gamma_3=g_3$), then
Eq.\ (\ref{rho_xx.Q1D.H>T}) produces a linear temperature dependence for
resistivity in agreement with Lebed's results \cite{Lebed89a,Lebed94}
for $T\ll E_H$. When renormalization of $\gamma_3$ is taken into
account, $\gamma_3({\bf q},0)$ diverges at a certain wave vector
$q_x=NG$ and $q_y=Q_y$ as $T\to T_c$. Assuming that only the term with
$q_x=NG$ dominates the sum (\ref{rho_xx.Q1D.H>T}), we find:
\begin{equation}
   \rho_{xx}^{(2D)} \approx \frac{\pi^2 b^2c T}{e^2 v_F}
   \int \, dq_y |{\tilde \gamma}_3(NG,q_y,0)|^2
   \sum_{n=-\infty}^{\infty} A_{n}(q_y)A_{N-n}(-q_y).
\label{rho_xx.NG}
\end{equation}
It is important to emphasize that the function $F({\bf
  q},\varepsilon)$ (\ref{F:H}), while restricting integration over
$\varepsilon$ and $q_x$, does not restricted integration over $q_y$
significantly (unlike in zero magnetic field, Sec.\ \ref{sec:H=0}),
because the Fourier coefficients $A_n(q_y)$ are nonsingular,
temperature-independent functions of $q_y$.  Thus, the integrals over
$q_y$ in Eqs.\ (\ref{rho_xx.Q1D.H})--(\ref{rho_xx.NG}) are not
restricted to the vicinity of the FISDW vector $Q_y$. This does not
allow us to replace $\gamma_3(NG,q_y,0)$ by $\gamma_3(NG,Q_y,0)$ and
take the latter outside of the integral, like in Eq.\ 
(\ref{rho_xx.T0}). The integration over $q_y$ reduces the divergence
of Eq.\ (\ref{rho_xx.NG}) at $T\to T_c$ and makes the resistivity
precursor effect of SDW in a magnetic field weaker than without the
field. This happens because the phase-space restrictions discussed in
Sec.\ \ref{sec:H=0} are consequences of the 2D nature of the electron
dispersion law (\ref{dispersion}) at zero magnetic field, whereas a
nonzero field makes the electron spectrum (\ref{dispersion1D})
one-dimensional.

In Fig.\ \ref{fig:rxxH}, we show temperature dependences of
resistivity calculated via Eqs.\ (\ref{rho_xx.Q1D}) and (\ref{F:H})
for $t_b'=20$ K, the phases $\varphi_b=0$, $\pi/27$, and $\pi/4$, and
the magnetic fields $H=0$, 5, 15, and 25 T. In the case of
$\varphi_b=0$ (the top panel), only a very strong magnetic field
$H=25$ T restores the negative slope of resistivity,
$d\rho_{xx}/dT<0$. This result cannot be applied to explaining
experiments \cite{Jerome95,Jerome96b,Chaikin98}, because for
$\varphi_b=0$ the slope of resistivity is already negative at
$15<T<60$ K in zero magnetic field, which does not agree with the
experiment. On the other hand, once we increase $\varphi_b$ to make
the zero-field slope positive, the effect of the magnetic field
becomes very weak outside of a narrow vicinity of $T_c$ for both a
very small phase $\varphi_b=\pi/27$ (the middle panel of Fig.\ 
\ref{fig:rxxH}) and a rather big phase $\pi/4$ (the bottom panel of
Fig.\ \ref{fig:rxxH}). If we keep $\varphi_b=0$ and increase $t_b'$ to
the values 30 K or 40 K, the zero-field slope becomes positive, but
only an enormous magnetic field of 50 T makes the slope negative (see
Fig.\ \ref{fig:rxxH30/40}) \cite{t_b'}.  One may conclude that in the
case $\varphi_b=0$ the negative slope of resistance occupies a
substantial range of magnetic fields only at very strong fields such
that $E_H>t_b'$: $H=25$ T for $t_b'=20$ K (the top panel of Fig.\ 
\ref{fig:rxxH}) and $H=50$ T for $t_b'=30$ and 40 K (Fig.\ 
\ref{fig:rxxH30/40}). This scenario is hard to reconcile with the
experiment, because, according to the standard theory of FISDW
\cite{Montambaux85}, the condition $E_H>t_b'$ corresponds to the
magnetic fields where the last, $N=0$ phase transition in the FISDW
cascade takes place, whereas the negative slope in resistance occurs
at much lower magnetic fields.

We conclude that the FISDW precursor scenario \cite{Gorkov95,Gorkov96}
cannot explain the anomalous temperature dependence of resistivity in
Q1D conductors in a magnetic field observed in experiments
\cite{Jerome95,Jerome96b,Chaikin98}. The insulating temperature
dependence of resistivity, $d\rho_{xx}/dT<0$, does exist in this
scenario at $T\gg T_c$ for a certain choice of the transverse
dispersion law of electrons, however the effect is present even
without magnetic field in this case. If the transverse dispersion law
is modified to suppress the insulating behavior in zero field, then
applying a magnetic field produces a negative slope in the temperature
dependence of $\rho_{xx}(T)$ only either in a narrow vicinity of $T_c$
or at very strong fields such that $E_H>t_b'$.

\section{Temperature dependence of nuclear magnetic relaxation}
\label{sec:NMR}

In Ref.\ \cite{Jerome95}, temperature dependence of the NMR relaxation
rate in $\rm(TMTSF)_2ClO_4$ was measured and was discussed as an
evidence for a charge pseudogap formation. In this section, we study
the effect of a magnetic field on the NMR relaxation rate within the
FISDW precursor scenario.

According to the theory of the NMR relaxation via the electron spin
fluctuations \cite{Moriya63} (see also Ref.\ \cite{Jerome94}), the NMR
relaxation rate $1/T_1$ is proportional to the imaginary part of the
electron spin susceptibility:
\begin{equation}
  \frac{1}{T_1} \propto T \sum_{\bf q} \frac{{\rm Im} \chi({\bf
      q},\omega_n)}{\omega_n},
\label{T1Moriya}
\end{equation} 
where $\omega_n$ is the nuclear Larmor frequency. The coefficient of
proportionality in Eq.\ (\ref{T1Moriya}) depends on the hyperfine
interaction parameters. In Q1D compounds, the sum over the wave
vectors ${\bf q}$ in Eq.\ (\ref{T1Moriya}) can be separated into the
contribution from the uniform spin susceptibility at $q_x$ close to
$q_x=0$ and the contribution from the antiferromagnetic spin
fluctuations at $q_x$ close to $q_x=2k_F$:
\begin{equation}
  T_1^{-1} = T_1^{-1}(q_x\approx0)+T_1^{-1}(q_x\approx2k_F).
\label{T1Q1D}
\end{equation}
In Refs.\ \cite{Jerome95,Jerome96a}, the temperature dependences of
each term in Eq.\ (\ref{T1Q1D}) were measured separately in
$\rm(TMTSF)_2ClO_4$ in the magnetic field 15 T along the ${\bf c}^*$
axis.  It was found that the antiferromagnetic term
$T_1^{-1}(q_x\approx2k_F)$ starts to grow below the same temperature
$T_{\rm min}$ that separates the metallic and insulating temperature
dependences of resistivity ($d\rho_{xx}/dT >0$ and $d\rho_{xx}/dT
<0$). It was claimed that the growth of $T_1^{-1}(q_x\approx2k_F)$ is
a manifestation of the opening of a charge pseudogap in the spectrum of
electron excitations.

In the ladder approximation, the renormalized spin susceptibility
$\chi({\bf q},\omega)$ is given by the Feynman diagrams similar to
those shown in Figs.\ \ref{fig:DWloop} and \ref{fig:vertices}:
\begin{equation}
  \chi({\bf q},\omega)=
   \frac{\chi_0({\bf q},\omega)[1-g_2\chi_0(-{\bf q},-\omega)]}
  {[1-g_2\chi_0({\bf q},\omega)] [1-g_2\chi_0(-{\bf q},-\omega)]
   -g_3^2\chi_0({\bf q},\omega) \chi_0(-{\bf q},-\omega)}.
\label{chi:ren}
\end{equation}
Neglecting the umklapp splitting of the FISDW instability, i.~e.\ 
setting $g_3=0$ in the denominator of (\ref{chi:ren}), and taking the
zero-frequency limit in Eq.\ (\ref{T1Moriya}), because the nuclear
Larmor frequency is small compared to all other energies, we find:
\begin{equation}
   \frac{1}{T_1(q_x\approx2k_F)} \propto T \sum_{\bf q} 
   \frac{ \lim_{\omega \to 0} {\rm Im} \chi_0({\bf q},\omega)/\omega}
   {[1-g_2\chi_0({\bf q},0)]^2}.
\label{T1SDW}
\end{equation}
$T_1^{-1}(q_x\approx2k_F)$ given by Eq.\ (\ref{T1SDW}) does not depend
on the phase $\varphi_b$.  The temperature dependence of
$T_1^{-1}(q_x\approx2k_F)$ calculated via Eq.\ (\ref{T1SDW}) with
$t_b'=20$ K is shown in Fig.\ \ref{fig:NMRH}.  We see that
$T_1^{-1}(q_x\approx2k_F)$ behaves in a magnetic field in the same way
as resistivity does, i.~e.\ it grows and deviates from the zero-field
curve only in a narrow vicinity of the critical temperature. Thus, we
conclude that the FISDW precursor scenario does not agree with the
experimental behavior of the NMR relaxation rate in
$\rm(TMTSF)_2ClO_4$ in a strong magnetic field.  The results of our
calculations are very similar to those of Ref.\ \cite{Bourbonnais93}
for a SDW transition without magnetic field \cite{cutoff}.

\section{Conclusions}
\label{sec:crfl:Concl}

In this paper we presented a heuristic, semiphenomenological
explanation of the anomalous temperature dependence of resistivity of
Q1D conductors in a magnetic field observed in experiments
\cite{Jerome95,Jerome96b,Chaikin98} reviewed in Sec.\ \ref{sec:intro}.
According to this picture (Sec.\ \ref{sec:qualitative}), a Q1D
conductor behaves like an insulator ($d\rho_{xx}/dT<0$), when its
effective dimensionality is one, and like a metal ($d\rho_{xx}/dT>0$),
when its effective dimensionality is greater than one. Applying a
magnetic field reduces the effective dimensionality of the system and
switches the temperature dependence of resistivity between the
insulating and metallic laws depending on the magnitude and
orientation of the magnetic field. Using this picture, we predicted
that the Danner oscillations of $\rho_{cc}$ may be observed when a
magnetic field is rotated in the magic plane from the ${\bf b}+{\bf
  c}$ direction toward the {\bf a} direction. We critically analyzed
whether various microscopic models suggested in literature can justify
our heuristic picture and found that none of the models is fully
satisfactory (Sec.\ \ref{sec:review}). We studied the FISDW precursor
scenario suggested by Gor'kov \cite{Gorkov95,Gorkov96} in detail both
analytically and numerically (Sec.\ \ref{sec:Q1D}--\ref{sec:NMR}) and
found that it does not agree with the experimental observations. In
the rest of this section, we speculate about possible alternative
approaches to solving the problem.

1. Within the FISDW precursor scenario, we studied only the umklapp
scattering rate $1/\tau$. In the language of Feynman diagrams,
$1/\tau$ is related to the imaginary part of the electron self-energy
$\Sigma$.  Via the Kramers-Kronig relations for $\Sigma$, any
precursor effects in Im$\Sigma$ should also affect Re$\Sigma$. Because
Re$\Sigma$ is related to the electron density of states, a pseudogap
may open in the single-electron spectral density as a precursor of
FISDW. Potentially, the pseudogap may affect transport properties of
the system as strongly as the renormalization of scattering rate
studied in this paper.  Re$\Sigma$ also determines the residue $Z$ of
the single-electron Green function, which is finite for a Fermi liquid
and vanishes for a Luttinger liquid. Re$\Sigma$ can be
straightforwardly calculated using the methods of this paper. On the
other hand, relationship between the single-electron and transport
properties is not straightforward. For example, the residue $Z$ may
cancel out from resistivity due to the Ward identities
\cite{Nozieres}.  Reliable calculations of resistivity require taking
into account corrections to the vertex of interaction between
electrons and electromagnetic field, which is a difficult problem for
inelastic electron-electron interaction \cite{Mahan90}.

2. The ladder approximation utilized in this paper does not take into
account contributions from superconducting channel, which, on one
hand, plays a very important role in one dimension and, on the other
hand, is sensitive to a magnetic field.  The magnetic field may induce
an insulating behavior by suppressing superconducting
fluctuations, which compensate insulating fluctuations at zero
magnetic field.  Including both density-wave and superconducting
channels naturally leads to the parquet equations.  If the parquet
equations in a magnetic field
\cite{Yakovenko87b,Yakovenko87a,Yakovenko91a} are generalized to
include the umklapp amplitude, one may try to find out whether they
can be decomposed into two disconnected sets of equation analogously
to the spin-charge separation in 1D case. This is a nontrivial
possibility, because the number of coupled equations is infinite, and
the equations are nonlinear. Nevertheless, if the equations do
decouple, the two sets of equations would diverge at two different
temperatures, one of which could be identified with $T_{\rm min}$ and
another with $T_c$. An additional technical problem is that the
equations of Refs.\ \cite{Yakovenko87b,Yakovenko87a,Yakovenko91a} are
applicable only at $T\alt E_H$, whereas $T_{\rm min}\sim E_H$, so the
parquet equations need to be derived in the difficult range $T\agt
E_H$. The parquet approach may require to assume a rather small
effective value for $t_b$ \cite{Yakovenko85c,Yakovenko88g}.

3. Electron-electron scattering contributes to electrical resistivity
only via the umklapp processes, which do not conserve the total
electron momentum \cite{Ziman}. For commensurate systems, such as
$\rm(TMTSF)_2X$, the umklapp scattering process that changes the total
electron momentum by $4k_F=2\pi/a$, where $a$ is the lattice spacing
in the {\bf a} direction, is usually considered. However, in the
presence of a magnetic field, the total electron momentum may also
change by a multiple of $G$, the wave vector of the magnetic field.
Since this kind of umklapp exists only in a magnetic field, it may be
a natural source of magnetoresistance in a Q1D system. It is
reasonable to expect that this mechanism works effectively only at
$T\alt E_H$. This could generate $T_{\rm min}\sim E_H$ without
invoking pseudogaps originating from renormalization, whose values
tend to be close to $T_c$. On the other hand, this mechanism would not
explain the insulating behavior of the transverse resistivities
$\rho_{yy}$ and $\rho_{zz}$ and the NMR relaxation rate.  However, the
recent experiment \cite{Chashechkina}, where metallic temperature
dependence was found for $\rho_{zz}$ and insulating for $\rho_{xx}$
and $\rho_{yy}$, may be in favor of this or another kinetic mechanism
and against a charge-gap scenario.

Undoubtedly, the anomalous temperature dependence of resistivity in
Q1D conductors in a magnetic field poses a tough theoretical challenge
and solving this difficult puzzle would greatly enrich the
condensed-matter physics.

\section*{Acknowledgments}

We are grateful to P.~M.~Chaikin, N.~Dupuis, L.~P.~Gor'kov, and
D.~J\'{e}rome for useful discussions. We also thank L.~Balicas,
E.~Chashechkina, and J.~Moser for sharing their unpublished
experimental data with us.  This work was partially supported by the
NSF under grant DMR--9417451 and by the David and Lucile Packard
Foundation.

\section*{Appendix}

The absolute values of $\rho_{xx}$ in $\rm(TMTSF)_2PF_6$ can be found
in Figs.\ 1 and 2 of Ref.\ \cite{Jacobsen81}, Fig.\ 20 of Ref.\ 
\cite{Jerome94}, Fig.\ 4 of Ref.\ \cite{Jerome98}, and Figs.\ 2 and 3
of Ref.\ \cite{Gorkov97}. From these data, as well as from Ref.\ 
\cite{Jerome}, we find that $\rho_{xx} \approx 18\:\mu\Omega$~cm at
$T=20$ K.  Comparing this experimental value of resistivity with the
theoretical curves in Fig.\ \ref{fig:rxx}, we find ${\tilde g}_3=0.17$
at $\varphi_b=0$ or ${\tilde g}_3=0.77$ at $\varphi_b=\pi/4$. On the
other hand, taking the value $\rho_{xx}\approx1$ $m\Omega$~cm at
$T=300$ K \cite{Jacobsen81,Jerome94}, we would find ${\tilde
  g}_3\approx1$.

In Ref.\ \cite{Gorkov97}, the experimental temperature dependence of
$\rho_{xx}$ in $\rm(TMTSF)_2PF_6$ was fit neglecting renormalization
of $\gamma_3$, and the value $g_3^{(GM)}=0.21$ was found. However,
because our Eq.\ (\ref{rho:var.pr}) differs from Eq.\ (3) of Ref.\ 
\cite{Gorkov97} by a factor $(2\pi)^3$, and a factor 0.5 is missing in
Eq.\ (7) of Ref.\ \cite{Gorkov97}, the value $g_3^{(GM)}=0.21$
corresponds to ${\tilde g}_3 =2\pi^{3/2}g_3^{(GM)} = 2.34$ in our
notation. The ratio of this value and the values quoted in the
preceding paragraph, $2.34/0.77=3.3$, represents the effect of
renormalization of $\gamma_3$ by the SDW diagrams.

According to the Drude formula, resistivity along the chains is
\begin{equation}
  \rho_{xx}=\frac{\pi bc\hbar}{2e^2
  v_F\tau}=\frac{4\pi}{\omega_p^2\tau},
\label{omega_p}
\end{equation}
where $\tau$ is the relaxation rate, and $\omega_p^2=8e^2 v_F/bc
\hbar$ is the plasma frequency. Using the values of $v_F$, $b$ and
$c$, we find $\omega_p=970$ cm$^{-1}$, which is close to the value
$\omega_p=1.1 \times 10^4$ cm$^{-1}$ from Ref.\ \cite{Gorkov97}.
Comparing Eq.\ (\ref{omega_p}) with the quoted above experimental
values of $\rho_{xx}$, we find $1/\tau \approx 40$ K at $T=20$ K and
$1/\tau\approx 2200$ K at $T=300$ K. One may notice that both of these
values for $1/\tau$ are greater than the corresponding temperatures:
$1/\tau>T$. For that reason, it may be more appropriate to use
$1/\tau$, rather than $T$, as an infrared cutoff of the
renormalization.



\begin{figure}
\caption{Feynman diagram for the bare density-wave susceptibility
  $\chi_0(x'-x,{\bf q}_{\perp},i\Omega_m)$. The solid and dashed lines
  represent the Green functions of the $+$ and $-$ electrons.}
\label{fig:DWloop}
\end{figure}

\begin{figure}
\caption{Feynman diagrams for the renormalized vertices of forward
  ($\gamma_2$) and umklapp ($\gamma_3$) scattering in the ladder
  approximation. The wavy lines represent the bare, unrenormalized
  vertices of forward ($g_2$) and umklapp ($g_3$) scattering, whereas
  the circles represent the corresponding renormalized vertices. The +
  and $-$ electron lines (the solid and dashed lines) are implied to
  have opposite spins. Spin is conserved along the thin lines inside
  the circles.}
\label{fig:vertices}
\end{figure}

\begin{figure}
\caption{Phase diagram of a Q1D conductor in the temperature (T) vs
  magnetic field (H) plane calculated from Eq.\ 
  (\protect\ref{condFISDW}) with neglected umklapp interaction
  $g_3=0$.  Various symbols denote the integer values $N$ of the
  quantized wave vector of FISDW, $q_x=NG$.}
\label{fig:PhaseDiagram}
\end{figure}

\begin{figure}
\caption{Temperature dependences of longitudinal resistivity normalized
  to the dimensionless umklapp scattering amplitude,
  $\rho_{xx}(T)/\tilde{g}_3^2$, at zero magnetic field for different
  values of the phase $\varphi_b$. The solid lines correspond to
  $t_b'=0$, in which case resistivity diverges at $T\to T_0$, where
  $T_0=14.7$ K is the SDW transition temperature. The dotted lines
  correspond to $t_b'=20$ K, in which case there is no SDW transition
  at zero magnetic field.}
\label{fig:rxx}
\end{figure}

\begin{figure}
\caption{Temperature dependences of the umklapp resistivity at $t_b'=20$
  K for different values of the phase $\varphi_b$ in the magnetic
  fields 5, 15, and 25 T (curves a, b, and c, respectively) and
  without magnetic field (dots).}
\label{fig:rxxH}
\end{figure}

\begin{figure}
\caption{Temperature dependences of the umklapp resistivity at $t_b'=30$
  K (top panel) and $t_b'=40$ K (bottom panel) for $\varphi_b=0$ in
  the magnetic fields 5, 25, and 50 T (curves a, b, and c,
  respectively) and without magnetic field (dots).}
\label{fig:rxxH30/40}
\end{figure}

\begin{figure}
\caption{Temperature dependences of the $2k_F$ component of the 
  NMR relaxation rate (in arbitrary units) at zero (dots) and nonzero
  (solid curves) magnetic field for $t_b'=20$ K.}
\label{fig:NMRH}
\end{figure}


\begin{thebibliography}{10}
\bibitem{Jerome95-note} The results shown in Fig.\ 1 of Ref.\ 
  \cite{Jerome95} are somewhat spoiled by the fact that the two sets
  of measurements were performed in two different laboratories, which
  resulted in different cooling rates of the sample and a big gap
  between the curves corresponding to the very close fields $H=12.3$ T
  and 12.5 T.
\bibitem{cutoff} Technically, the high-energy cutoff $\hbar v_F/\pi
x_0\sim\varepsilon_F\sim t_a$ that we use for the calculation $\chi_0$
is not quite appropriate, because at the energies $\varepsilon\geq
t_b$ the system is truly one-dimensional, and the Peierls instability
is cancelled by the Cooper one in the parquet approximation
\cite{Yakovenko87b}.  So the more appropriate energy cutoff would be
$t_b$, which is much smaller than $\hbar v_F/\pi x_0$.  Nevertheless,
it appears that changing the cutoff would only affect the numbers and
would not alter the qualitative conclusions of the paper.
\bibitem{resistivity} By 1D or 2D case we mean that $t_b=0$ or
  $t_c=0$; that is, either the chains or the plains of a Q1D conductor
  are decoupled. However, in all cases, the expressions for
  $\rho_{xx}$ in this paper have the dimensionality of 3D resistivity.
  Resistivity of a single chain would be given by Eq.\ 
  (\ref{rho_xx.1D}) divided by $bc$.
\bibitem{b} More precisely, the expressions for $G$ and $E_H$ should
  contain not the lattice spacing $b$, but its projection
  perpendicular to the {\bf a} axis; however, the difference is not
  significant.
\end{thebibliography}
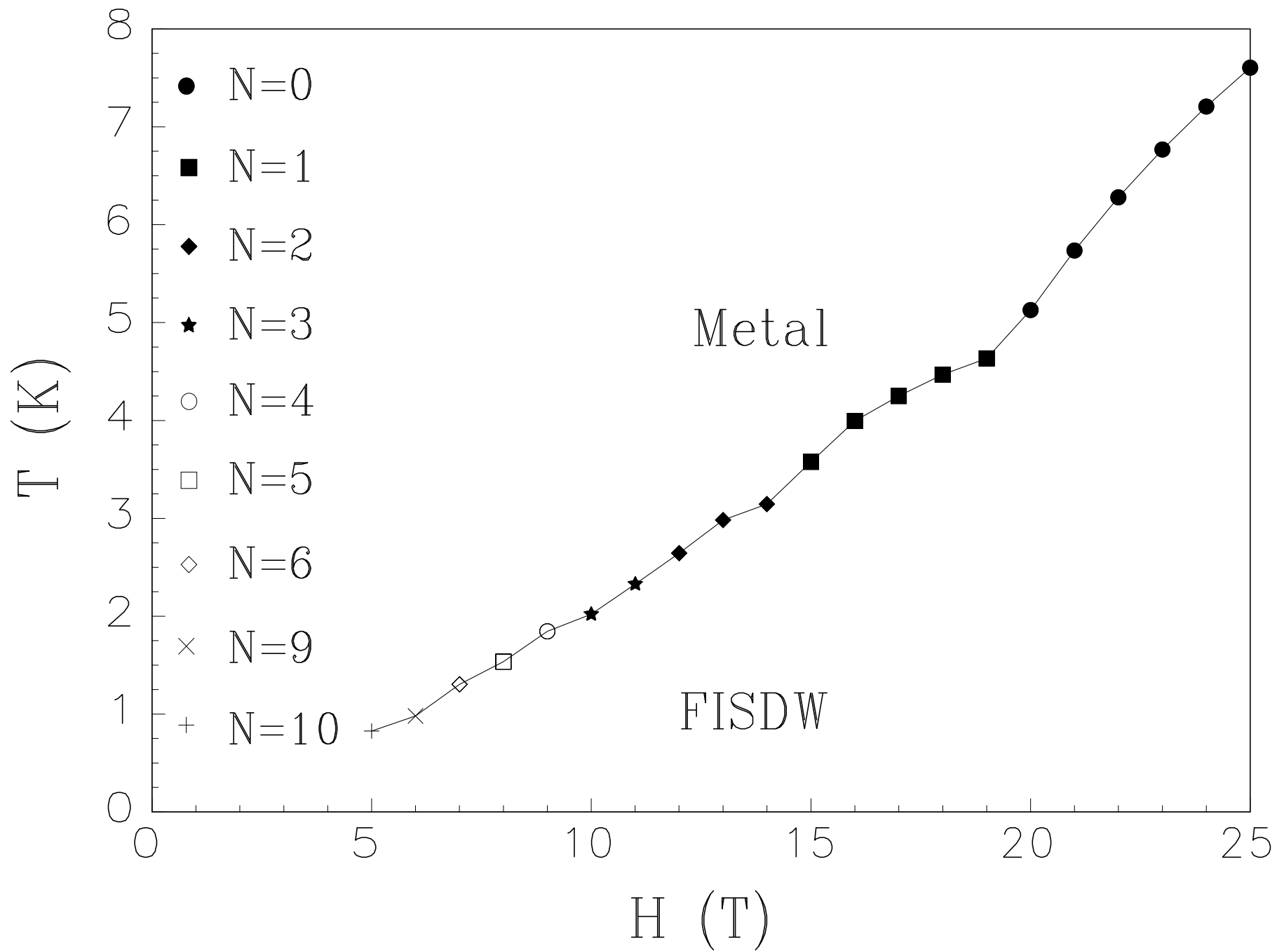
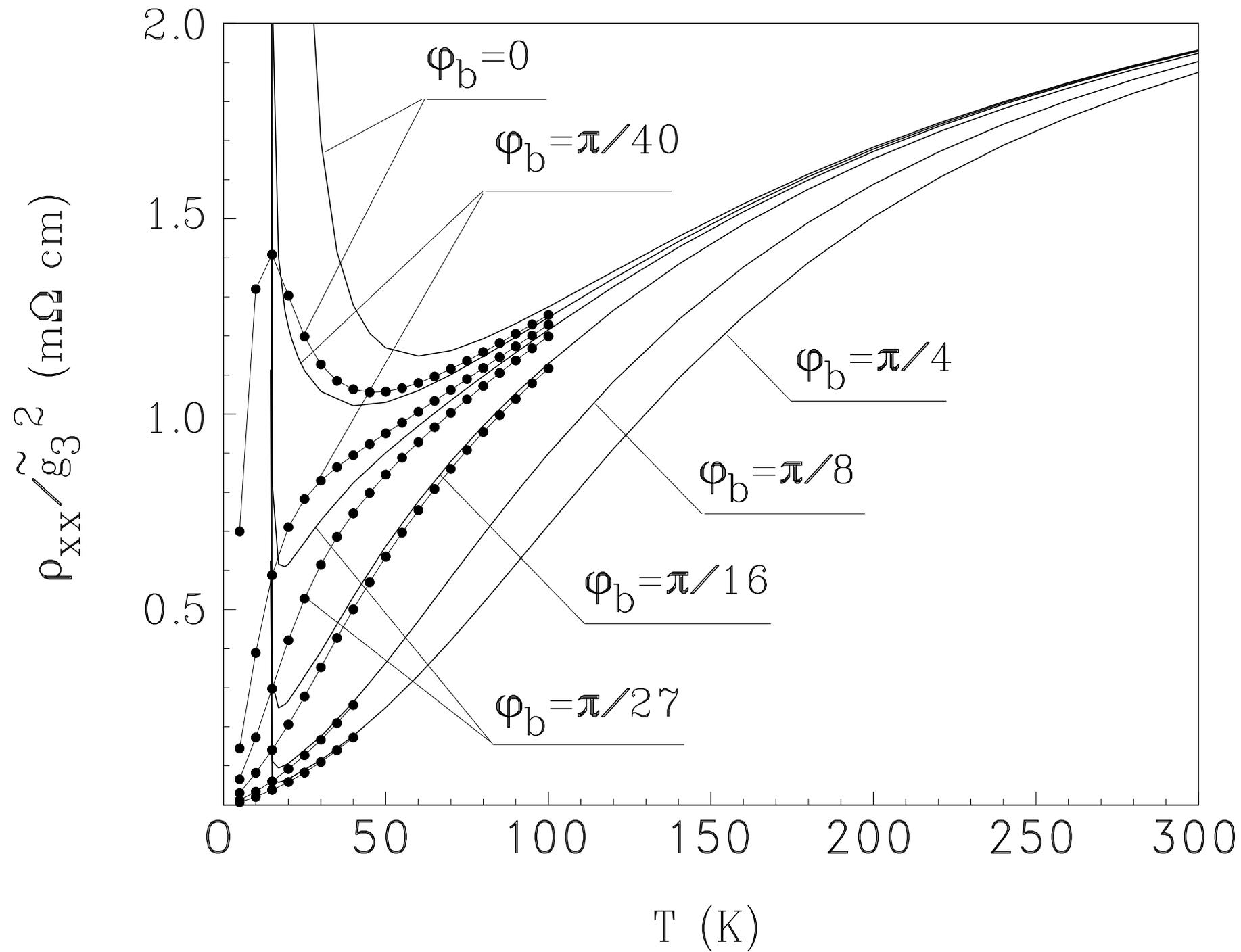
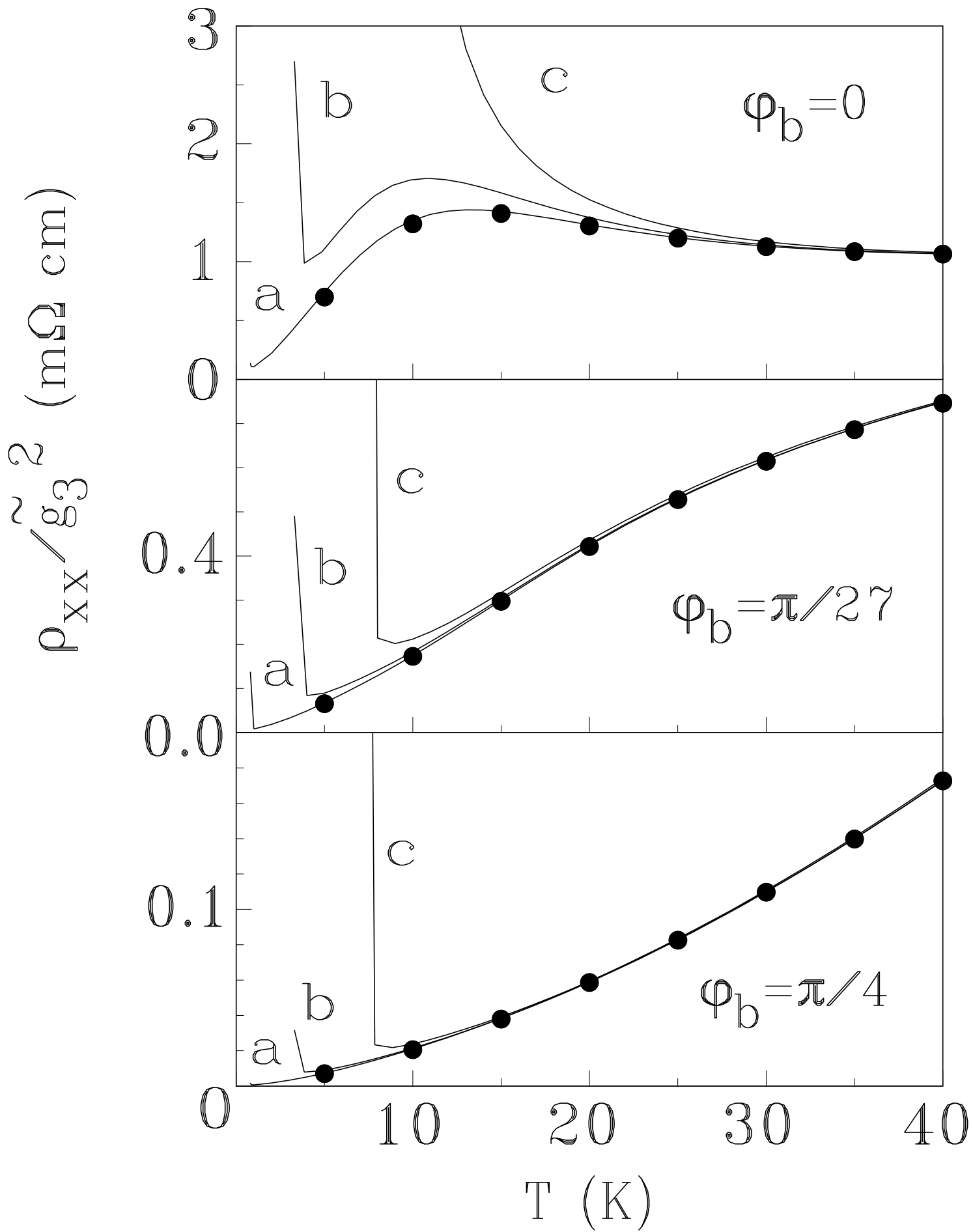
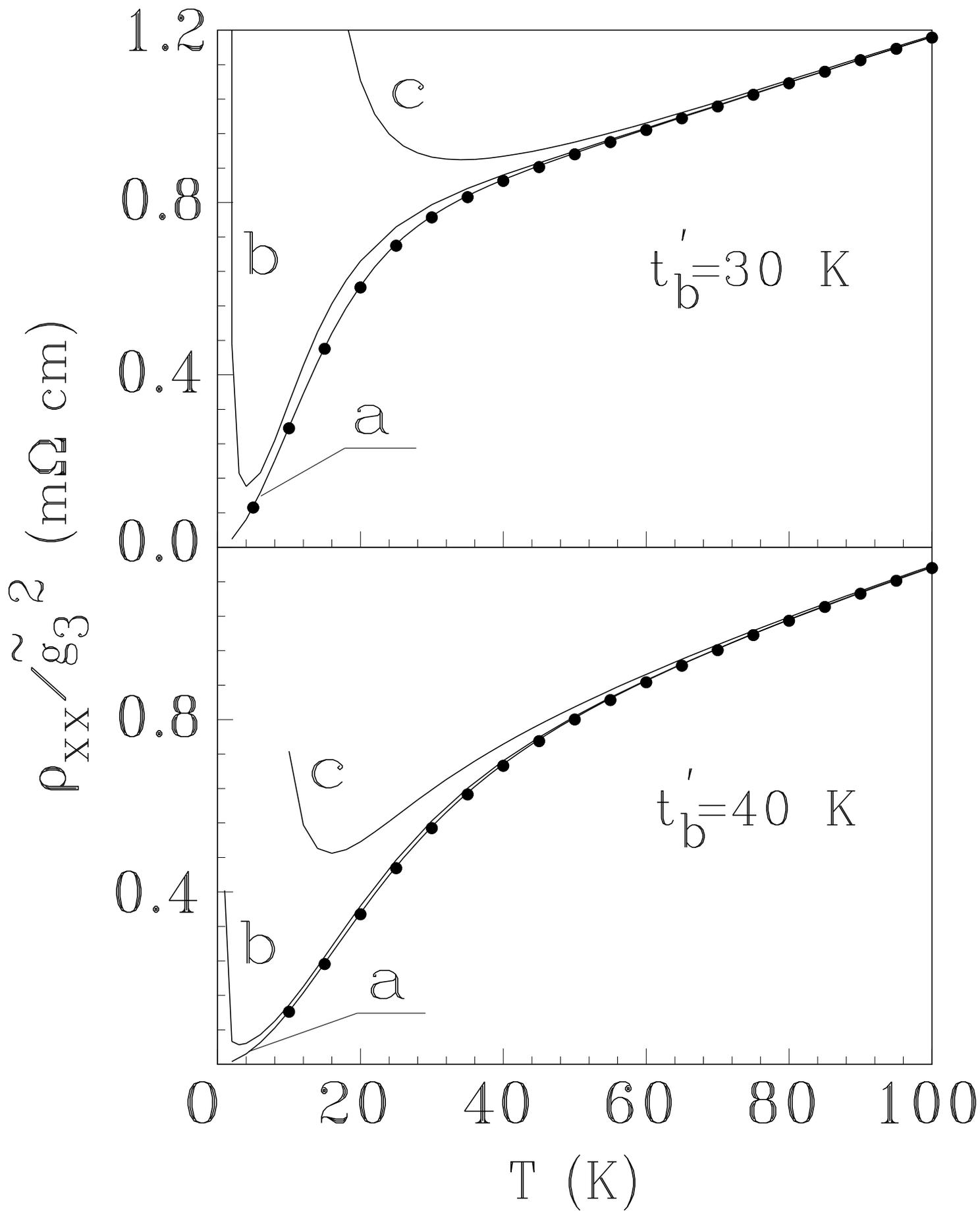
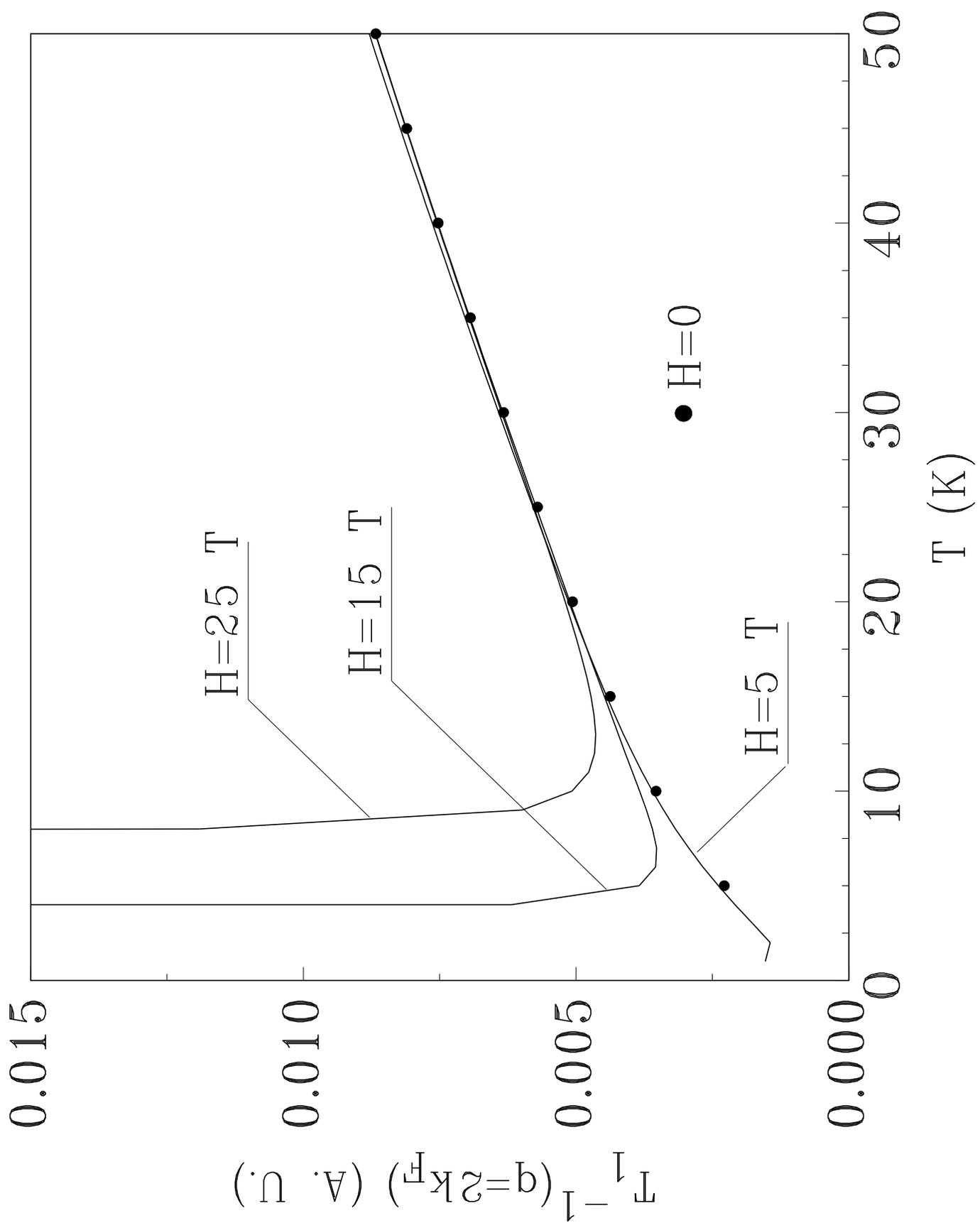
\end{document}